\documentclass[12pt]{article}

\newcommand{\bbr}{I\!\! R}

\newcommand{\m}{\mathrm}
\newcommand{\be}{\begin{equation}}
\newcommand{\ee}{\end{equation}}
\newcommand{\ba}{\begin{eqnarray}}
\newcommand{\ea}{\end{eqnarray}}

\usepackage{graphicx}
\usepackage{amssymb}
\usepackage{amsmath}
\usepackage[T1]{fontenc} 
\usepackage[ansinew]{inputenc} 
\usepackage[nosort]{cite}
\newcommand{\inbar}{\vrule height1.57ex width.4pt depth0pt}
\newcommand{\SW}{\relax{\hbox{$\ \inbar\kern-.285em{\rm S}$}}}

\begin{document}
\thispagestyle{empty}
\begin{center}

\null \vskip-1truecm \vskip2truecm

{\Large{\bf \textsf{Intrinsic Torsion, Extrinsic Torsion, }}}
\vskip0.5truecm

{\large{\bf \textsf{and the}}}
\vskip0.5truecm

{\Large{\bf \textsf{Hubble Parameter}}}

\vskip1truecm

{\large \textsf{Brett McInnes}}

\vskip1truecm

\textsf{\\  National
  University of Singapore}

\textsf{email: matmcinn@nus.edu.sg}\\

\end{center}
\vskip1truecm \centerline{\textsf{ABSTRACT}} \baselineskip=15pt
\medskip

We study the intrinsic and extrinsic torsions (defined by analogy with the intrinsic and extrinsic curvatures) of the spatial sections of torsional spacetimes. We consider two possibilities. First, that the intrinsic torsion might prove to be directly observable. Second, that it is not observable, having been ``inflated away'' in the early Universe. We argue that, even in this second case, the extrinsic torsion may grow during the inflationary era and be non-negligible at reheating and thereafter. Even if the spatial intrinsic curvature and torsion are too small to be detected directly, then, the extrinsic torsion might not be. We point out that, if its presence is not recognised, the extrinsic torsion could lead to anomalies in the theoretical estimate of the Hubble parameter ---$\,$ a result with obvious potential applications. We stress that extrinsic torsion is by far the most natural way to produce such anomalies, simply because it mixes naturally with the Hubble parameter; that is, the second fundamental form of a spacelike section depends on a sum of two terms, one determined by the Hubble parameter, the other by the extrinsic torsion.

\newpage
\addtocounter{section}{1}
\section* {\large{\textsf{1. In the Footsteps of C.F. Gauss}}}
In 1821, (Johann) Carl Friedrich Gauss began a new phase of his extraordinarily eclectic scientific career: he oversaw a triangulation of a part of the Kingdom of Hannover \cite{kn:breit}. Gauss himself took a direct hand in this endeavour, going into the field and actually \emph{measuring} (with the aid of a device, the ``heliotrope'', of his own invention\footnote{See the Wikipedia article for Heliotrope (instrument).}) the side lengths and angles of 26 triangles. These fitted into a large triangle (with side lengths 69, 84, and 106 kilometres), defined by the peaks of three mountains, the Hoher Hagen, Brocken, and the Grosser Inselberg, and Gauss determined the angles of this triangle with exquisite precision and accuracy.

Gauss' motives for performing the measurements of his largest triangle have been a subject of a lively debate (see for example \cite{kn:beich,kn:marder1,kn:marder2}), to which we cannot do justice here. His main objective was undoubtedly to check for errors in his previous work, and to detect deviations from exact sphericity of the shape of the Earth. But it has often been suggested that, in addition to this, Gauss was looking for something more fundamental: a deviation of three-dimensional space from being exactly Euclidean\footnote{We use ``Euclidean'' (meaning it in the sense that Gauss would have understood) to avoid the word ``flat'', for reasons which will appear.}. Of course, he did not find it; the sum of the angles of his large triangle was found to be 180 degrees, with an astonishing accuracy of within two thirds of a second of arc.

Gauss, the great astronomer, would have been well aware that any deviation from Euclidean geometry he could detect in this manner would probably long since have been obvious in the skies (and in fact Schwarzschild considered this issue in numerical detail, just prior to the advent of General Relativity \cite{kn:sch1}, translated as \cite{kn:sch2}). But it would be in keeping with his characteristic correctness in matters of scientific integrity to look for such an effect, even though he did not expect to find one.

Let us leave the historical debate to one side, and entertain the possibility that Gauss was indeed interested, if only casually, in studying the geometry of the three-dimensional space we inhabit. We wish to draw the reader's attention to a simple fact, for which there is abundant documentary evidence (see again \cite{kn:breit,kn:beich,kn:marder1,kn:marder2}): Gauss believed that the geometry of space could be determined by \emph{observations}: it is an empirical matter. Straight lines are those lines that one \emph{observes directly} to deviate neither to left nor to right; and one could go out with his (sufficiently advanced) geodetic instruments, and use them to \emph{observe directly} what these straight lines do, concluding that space is not Euclidean, if that should be the case. This was a controversial position to take in Gauss' time, and indeed for a surprisingly lengthy period after him.

In modern language, what Gauss was (allegedly) looking for was evidence of \emph{geodesic deviation} in three-dimensional space. This is the basic object that distinguishes Euclidean from non-Euclidean geometry, and it is at the centre of our concerns here. Geodesic deviation is a \emph{non-linear rate of spreading} of a family of straight lines emanating from a point.

Notice that, in principle, Gauss might have observed this phenomenon of non-linear spreading near any point where two straight lines crossed; he formulated his results in terms of angle sums of triangles for historical and practical reasons. That is, the phenomenon of geodesic deviation is more fundamental than that of the departure of angle sums of triangles from 180 degrees. (In differential-geometric language, it depends on the linear connection, not on the metric as the concept of angle does.)

Now let us try to relate this discussion to a modern view of \emph{spacetime}. The latter does \emph{in fact} have a non-trivial geometric structure: this is of course revealed by cosmological and gravitational wave observations (see for example \cite{kn:ni,kn:gravwav,kn:valer}), which is one reason for the power of these observations to eliminate some less thoroughly geometric theories of gravitation \cite{kn:zuma}.

It follows that \emph{Gauss was right} if he suspected that the actual three-dimensional space of our experience exhibits non-zero geodesic deviation. The spatial sections defined \cite{kn:folia} by a given family of observers (including the one to which Gauss belonged) will inherit, generically, a non-trivial geometry: think, for example, of the spatial sections of the exterior Schwarzschild spacetime, associated with the distinguished observers for whom the spacetime is static. In other words, the geometry of spacelike sections is determined by the spacetime geometry, once the observers are specified; and that geometry is normally\footnote{In some cases, one can find a family of observers such that the component of spacetime curvature evaluated on the corresponding spatial sections is ``cancelled'' by the extrinsic curvature (the terms involving $\alpha$ in equation (\ref{FFFF}), below) of the foliation, leading to Euclidean spatial sections; this is of course the case for spatially Euclidean FRW cosmological models. However, this is a \emph{very unusual} arrangement, one that calls for explanation. The theory of Inflation provides that explanation, as we will discuss later in detail.} non-Euclidean.

In short: the effect Gauss was supposedly seeking is in fact there, on the local (though \emph{perhaps} not cosmological) level.

Our motivation for reminding the reader of Gaussian history is as follows. Recently, the concept of \emph{spacetime torsion} has attracted a great deal of attention: see \cite{kn:blaghehl} for a guide to earlier work, and \cite{kn:telebook,kn:mavro} for recent reviews. Many of these theories have proved to be untenable \cite{kn:ong} (but see also \cite{kn:golov}). But, more recently, several new and very interesting applications of torsion have arisen.

For example, spacetime torsion might be associated with dark matter \cite{kn:dark,kn:darker} and might in that way eventually be observable in gravitational wave observations, through its effects on the wave amplitudes and polarizations \cite{kn:polar}. There are reasons \cite{kn:eliz} to doubt whether such an effect could ever be large enough to be observable, but there may be a loophole involving primordial gravitational waves. Another very promising approach is to study the effects of torsion on the decay of binary orbits \cite{kn:bat}. Again, it has been suggested \cite{kn:salvio1,kn:kyrie,kn:gialamas,kn:salvio2,kn:salvio3} that torsion\footnote{Actually, following \cite{kn:percacci}, these works allow non-metricity as well as torsion.} arises very naturally in Inflationary cosmology: it may even allow us to address the question of a UV completion for Inflation. These recent developments, and others, justify a renewed interest in spacetime torsion.

Thus, the question as to whether the three-dimensional space of our experience has Euclidean geometry (and topology) remains today as pertinent as it was in Gauss' time. In the past, the geometry of the spatial sections of cosmological spacetimes could only be probed relatively crudely, through the contribution of the spatial curvature (meaning just the term $\Omega_k/a^2$, not the full tensor) to the Friedmann equation\footnote{But \emph{some} basic direct observations of spatial geometry have of course been made: most importantly, the observations confirming \cite{kn:hut} \emph{spatial isotropy} are of this kind.}. Now, however, gravitational wave theory and observations once again open the way to far more refined investigations of spatial geometry on cosmological scales: see for example \cite{kn:kumar} and its references.

In these works, one considers perturbations around background FRW spacetimes with spatial sections not assumed to be Euclidean, studying them using the curved-space Laplace equation, with the associated generalised scalar and tensor harmonics \cite{kn:harmonic}. This can in principle reveal any spatial curvature that may be present. It is no doubt possible, though perhaps challenging, to extend this work to include spatial torsion; if that can be done, the way is open to use it to explore the possibility that the spatial sections of our Universe have a geometry described by a small but detectable torsion tensor. The key point here is that this procedure probes the spatial geometry \emph{directly} (that is, not by deducing it from a knowledge of the ambient spacetime geometry).

In short, such work follows in Gauss' footsteps: it seeks to explore the intrinsic geometry of three-dimensional space \emph{directly and in detail}. In the foreseeable future, then, observational data on the details of that geometry may well become available, and it is conceivable that a non-zero intrinsic torsion might be found in that manner. (Work on the \emph{topology} of space has also reached a high level of sophistication: see for example \cite{kn:compact}.)

Thus far we have considered only what we might call the \emph{intrinsic} torsion of three-dimensional space, the torsion it has without regard to its embedding in spacetime. However, particularly in cosmological spacetimes, the extrinsic curvature of three-dimensional space plays a vital role, and one might well expect that ``extrinsic torsion'' should be important.

We need answers to the following questions. In a spacetime with non-zero torsion, how does one study the associated purely \emph{spatial}, intrinsic torsion of the spacelike sections corresponding to a distinguished family of observers? Next, if we can do this, how can we predict geodesic deviations in space, in the hope that these can be compared with the observations? Finally, can we evaluate the extrinsic torsion, and can we determine its physical meaning?

We answer the first two questions with a straightforward extension of classical submanifold theory, giving explicit expressions for intrinsic and extrinsic torsion. We use this theory to illustrate the fact that, while torsion is in many ways similar to curvature (indeed, there is a sense in which it \emph{is} a kind of curvature), there are also fundamental, and sometimes surprising, differences. Torsion can behave in ways that curvature in non-torsional theories cannot.

Our second application for these methods is in the theory of \emph{cosmic Inflation} \cite{kn:inflationaris}: a theory in which, of course, the geometry of three-dimensional space is a major theme. Inflation in GR would have it that space should be, to an excellent approximation, Euclidean. A less often emphasised role for Inflation is as an explanation of the fact that the Hubble parameter is large enough to be easily observed at the present time: this is due to the fact that Inflation increases the extrinsic curvature of the spatial sections of the early Universe, just as it decreases the intrinsic curvature.

We investigate inflationary theories incorporating spacetime torsion from this intrinsic/extrinsic point of view, finding, as expected, that the extrinsic torsion plays an essential role. In particular, the extrinsic torsion can have an effect on theoretical predictions of the Hubble parameter, a result which might well reward further attention.

The intention here is to obtain results of maximal generality, which can be applied in principle to most of the many torsional extensions of GR. We will, however, impose one major restriction: we do not allow ``\emph{non-metricity}'' ---$\,$ the metric is always taken to be compatible with the linear connection. Our principal reason for doing this is pragmatic: submanifold theory, our main mathematical technique here, is very greatly complicated by dropping compatibility. For example, the second fundamental form loses its usual geometric interpretation in that case, and the Gauss submanifold curvature formula is difficult to handle and to interpret.

We also wish to stress, however, that non-metricity is, from almost any point of view, radically different from torsion. If we do not insist on compatibility, then lengths of spacetime vectors cannot be compared remotely: they become globally ambiguous. But particle masses are nothing but the lengths of four-momenta, so in such a spacetime one would need to re-think the crucial classification of particles in terms of representations of the Poincar\'{e} group. As is well known \cite{kn:blaghehl}, this group is closely connected with torsion (see Appendix 2 of this work), so the distinction between the two sets of ideas is sharp\footnote{This is part of a, firmly negative, answer to the question raised in the title of \cite{kn:trinity}.}. In a related development, it has in fact been argued recently that the natural physical interpretation of ``non-metricity'' is in terms of violation of local Lorentz invariance \cite{kn:obhe}. Such violations have been constrained exceedingly severely by observations \cite{kn:stringent}.

None of this can be said of torsion. While we do not rule out a role for non-metricity, enough has been said to justify not considering it further in the present work.

\addtocounter{section}{1}
\section* {\large{\textsf{2. Spatial Geodesic Deviation in GR}}}
For the sake of clarity, let us first discuss the situation in the context of zero torsion, that is, using ordinary General Relativity.

Let us join Gauss in 1821 in rural Hannover. In imagination, let us suppose that he examines two intersecting straight lines in three-dimensional space (that is, not attached to the surface of the Earth), and is delighted to find that they spread apart non-linearly as he proceeds along them away from the intersection point. He believes that this implies that the angle sums of his triangles will almost certainly not be exactly 180 degrees, and he will proceed to check his result by confirming this. But that is incidental. The essential point is the non-linear spreading: Gauss has directly measured a non-zero \emph{geodesic deviation} in three-dimensional space.

Sadly, this is not what happened. But it could have.

Notice carefully that Gauss was not studying the trajectories of particles in \emph{spacetime}: his straight lines were straight lines in \emph{space}. This is of course a much less ambitious project, and that will allow us to avoid unnecessary complications later.

In geometric language: we suppose that the three-dimensional space of our (and Gauss') experience is a manifold endowed with a metric and with a metrically compatible linear connection. Consider a (canonically parametrised) family of autoparallels\footnote{We use this terminology, a little pedantically it may be, as a way of avoiding (as far as is reasonable) the word ``geodesic'', which is too closely associated with worldlines of free particles in \emph{four} dimensions.} (that is, ``straight lines'') with respect to this connection, take a particular autoparallel with tangent vector field $u$, and let $U$ be the displacement vector to other members of the family. Then, in our fantasy, Gauss' careful geodesy amounts essentially to a measurement of a non-zero value for $\nabla_u \nabla_u U$, where $\nabla$ is the derivative operator defined by the connection.

Now usually we would talk about this sensational discovery (had it actually happened) in a different way, as follows. It turns out (see \cite{kn:kobnom2}, Chapter VIII, whose notation we follow throughout this work) that, in Riemannian (and semi-Riemannian) geometry, $\nabla_u \nabla_u U$ can be expressed more conveniently in terms of a tensor, $R$: we have Jacobi's equation
\begin{equation}\label{A}
\nabla_u \nabla_u U \;=\; -\,R(U, u)u.
\end{equation}
Of course, $R$ is the \emph{curvature tensor} (of this three-dimensional Riemannian space). If Gauss had succeeded in observing a non-zero spatial geodesic deviation, we would discuss this (in the absence of torsion, that is, in the context of GR) by saying that he had observed non-zero spatial curvature. However, it is worth stressing that equation (\ref{A}) is just a mathematical convenience: the real object of interest here is the geodesic deviation; in this context, the curvature is just a useful measure of that object. (Of course, curvature can also be interpreted in terms of the path-dependence of parallel transport, but in this discussion we think of that as a derived consequence of the existence of geodesic deviation.)

Curvature really comes into its own if we imagine that we have a physical theory \emph{capable of predicting its value}. For then we can substitute that predicted value into the right side of equation (\ref{A}), solve for $U$, compute $\nabla_u\nabla_u U$, and confirm that it agrees with the value that Gauss has \emph{measured}.

In more detail: we can regard equation (\ref{A}) as a linear second-order differential equation for $U$. Using a dash to indicate a derivative with respect to the canonical parameter along a given autoparallel, we have
\begin{equation}\label{B}
U'' \;+\; R(U, u)u \;=\; 0.
\end{equation}
This can be solved explicitly, once initial conditions are given, in terms of Green's functions and so on: see Chapter 1 Section 6, and Chapter 2 Section 3, of \cite{kn:synge}. The standard theorems on the existence and uniqueness of second-order linear differential equations assure us that $U$ can be found and is unique. Once we have $U$, we can compute $U''$, and make the comparison with Gauss' observations. Thus, if we have a way of predicting the value of the three-dimensional curvature tensor, then we have a \emph{falsifiable physical theory} of three-dimensional space.

We, unlike Gauss, can carry through this programme, because we have a theory which does predict the value of the curvature tensor: General Relativity.

In fact, there is a technical complication here: GR actually gives us the curvature tensor of spacetime, not of space. To deal with this, we choose a distinguished family of timelike curves (``observers'') and assume (\emph{throughout this work}) that they define a ``$1\,+\,3$'' decomposition of spacetime, that is, a foliation of spacetime by spacelike hypersurfaces. In the language of submanifold theory (which, in its traditional form, is of course due to none other than C.F. Gauss; see Chapter VII of \cite{kn:kobnom2} for the modern version), the hypersurface is described by the \emph{second fundamental form} $\alpha$. The \emph{Gauss equation} of submanifold theory then allows us, given the spacetime curvature and $\alpha$, to make a definite prediction as to the curvature tensor of three-dimensional space, and then we can proceed as above. (We will not discuss this explicitly here, because we will do that below in the case where torsion is present.)

To summarise: Gauss showed us that geodesic deviation in three-dimensional space is directly (intrinsically) observable. GR makes definite predictions as to what Gauss should have found; so Gauss' observations could in principle have falsified GR, had it been known then.

Notice that we have nowhere made any assumptions about the behaviour of free particles. Our ``straight lines'' are straight lines in space. In GR, which we have been using up to this point, the worldlines of free particles are in fact straight lines in four-dimensional spacetime. However, in the torsional theories we are about to consider, worldlines of free massive particles are still curves of extremal proper time (and those of photons are still null limits of such \cite{kn:santana}), as they are in GR, but they are usually\footnote{The exception is the case of ``axial'' torsion. (The torsion is said to be axial if, when it is interpreted with the aid of a metric as a $(0, 3)$ tensor, it is antisymmetric in all pairs of ``slots''.)} not straight (that is, not autoparallels). That complicates the investigation of torsional \emph{spacetime} geometry, but it does not affect our (strictly three-dimensional) discussion here. This is what we meant earlier by saying that a focus on space, rather than spacetime, simplifies these matters very considerably. (We will of course have to study the way in which the spatial sections are embedded in \emph{spacetime}, and that will prove to be very important. But there, too, the dynamics of free particles plays no role.)

So much for GR. We now want to explain that an almost identical programme is practicable when torsion is present (though the technical details are interestingly different). (See \cite{kn:lemos} and \cite{kn:raych} for relevant recent work.)

\addtocounter{section}{1}
\section* {\large{\textsf{3. Spatial Torsion and Spatial Geodesic Deviation}}}
We now study geodesic deviation in any spacetime theory with a metric and a compatible linear connection with possibly non-zero torsion. We first clarify our point of view on torsion, and then proceed to the relevant technicalities from submanifold theory.

\addtocounter{section}{1}
\subsection* {\large{\textsf{3.1. Torsional Geometry}}}
Let us be very clear about our intentions here. On any differentiable manifold, there exists an uncountable infinity of possible metric tensors and another infinity of linear connections. But if the manifold has a definite geometry ---$\,$ meaning that tangent vectors have well-defined lengths, and that certain curves are distinguished as being straight (but with the possibility of giving rise to non-zero geodesic deviations) ---$\,$ this geometry will select a specific metric tensor and a specific linear connection to represent it. (It cannot select two metrics or two linear connections: there is only one geometry.) At this point, \emph{all of the other candidate metric tensors and linear connections must be regarded as mathematical fictions}. Geometry defines a metric and a linear connection ---$\,$ not the other way around.

Thus, to take a particularly important example, if the actual linear connection of spacetime has non-zero torsion, then the usual Levi-Civita connection generated by the metric belongs to the fictitious multitude. It is of course still possible to compute (for example) the curvature tensor of this connection; but, when torsion is present, this ``Levi-Civita curvature tensor'' is not the one that appears in the equation of geodesic deviation and so it does not describe the \emph{actual} geometry.

\emph{However}, mathematical fictions are of course often very useful. It is often convenient, for example, to write the Friedmann equation using a sum of terms involving energy ``density'' parameters, including a term of the form $\Omega_k/a^2$, where $a$ is the scale factor and where $\Omega_k$ is the ``density'' parameter contributed by the spatial curvature (in the absence of torsion). Naturally, one does not really believe that spatial curvature gives rise to a real energy density, but it is useful to pretend that it does.

We can continue this pretence even in the presence of torsion: the curvature parameter $\Omega_k$ continues to represent a fictitious ``density'' term in the Friedmann equation \emph{arising from the Levi-Civita ``curvature'' only}, but there will also be a potentially observable ``density'' parameter $\Omega_T$ arising from the torsion, and it too can be included in the Friedmann equation (see \cite{kn:barrow2}). Observational evidence for a non-zero $\Omega_T$ would provide indirect reason to believe in non-zero torsion, exactly as evidence for a non-zero $\Omega_k$ would indicate non-zero spatial curvature: indirect, in the sense that geodesic deviations are not being measured.

Fictitious though these parameters are, all this is very useful in making the connection with observations: for example, it is $\Omega_k$ that is shown to be close to zero by a careful analysis of the data \cite{kn:efstath} (see \cite{kn:suhail,kn:bengal} for more recent data and forecasts). But this should not distract us from the fact that the Levi-Civita ``curvature'' is not the actual curvature, not the curvature that we hope to measure by (for example) analysing tensor harmonics describing gravitational waves propagating through spatial sections.

With this understood, let us proceed. In the presence of both curvature and torsion, the equation of geodesic deviation takes the form (see again \cite{kn:kobnom2}, Chapter VIII)
\begin{equation}\label{C}
\nabla_u \nabla_u U \;=\; -\,\nabla_u\left[T(U, u)\right] \;-\;R(U, u)u.
\end{equation}
Here $u$, $U$, and $R$ are as before, and $T$ is the torsion tensor. (Bear in mind that we want to apply this to three-dimensional space.) We see at once that \emph{torsion is explicitly connected with geodesic deviation}, just as curvature is. Indeed, if torsion is present, there will be geodesic deviation \emph{even if the curvature is zero}. (For this reason we eschew the use of the word ``flat'', preferring ``Euclidean'' when we wish to refer to a geometry with zero geodesic deviation.)

Let us expand on this last point, because it brings into focus the profound ways in which torsion brings something new to spacetime geometry. We briefly digress to explore the geometry of three-dimensional space in the absence of curvature.

Geodesic deviation is now governed, using the same notation as in the preceding Section, by the very simple equation
\begin{equation}\label{G}
X'' \;+\; T(X, u)' \;=\; 0;
\end{equation}
simple, because it is essentially a first-order equation:
\begin{equation}\label{H}
X' \;+\; T(X, u) \;=\; K,
\end{equation}
where $K$ is a vector which is constant along the reference autoparallel, determined of course by the initial value of $X'$.

Given the torsion, we see that equation (\ref{H}) is in fact a linear dynamical system along the reference autoparallel, with the (1,\,1) tensor $T_u = T(\cdot\,,\, u)$, referred to some basis, as the system matrix. The whole vast machinery of dynamical systems theory now becomes available to us.

In the simplest possible case, the torsion varies slowly along a reference autoparallel, so that $T_u$ is approximately constant. Generically, this matrix will be non-singular, and then we can define $Y = X - T_u^{-1}K,$ so that the equation of spatial geodesic deviation is just
\begin{equation}\label{I}
Y'\;=\;-\,T_uY.
\end{equation}
This is the standard form for a dynamical system, leading to a well-understood classification of the possibilities: one will have spiral sinks, saddle points, and so on.

In general, however, $T_u$ will not be constant, and the resulting dynamical system can be difficult to handle. However, it makes sense to focus initially on situations in which the spatial geometry is symmetrical. Symmetry groups in torsional geometries tend to be relatively small: see \cite{kn:krss,kn:coley0,kn:coley1,kn:coley2}; but they still have the power to constrain the form of $T_u$, and, in the cases with the largest symmetry groups, much can be said about it. Thus, in the case of spatial sections with symmetries, we can hope to arrive at a classification of the possible forms that spatial geodesic deviation can take.

It might also be possible to use specialised techniques to study particular cases. For example, one might wish to study purely torsional primordial gravitational waves (which, as mentioned earlier, may escape the strictures of \cite{kn:eliz}) propagating through a ``spatially teleparallel'' cosmological spacetime. Here the spatial geometry will have a periodic structure, and so $T_u$, while not constant, is likewise periodic. This case too is well-understood: the techniques of \emph{Floquet Theory} \cite{kn:teschl} allow a complete analysis, essentially by finding a time-dependent change of variable that reduces a periodic system matrix to a constant one.

There is much scope for interesting work here, but let us leave that to one side: the point is that all this looks very unfamiliar. There are torsional geometries that cannot be understood in terms of any kind of curvature.

To resume our discussion of the general case: equation (\ref{C}) is central to our concerns in this work. It gives us a language in which to report observations of geodesic deviation, the language of curvature \emph{and} torsion. One of its immediate implications is that, in a sense, ``\emph{torsion is a novel kind of curvature}''. There is a sense in which this statement is literally true; this will be explained in Appendix 2, below.

Equation (\ref{C}) can be written a little more transparently in the form
\begin{equation}\label{CC}
\nabla_u \nabla_u U \;+\;T(\nabla_u U, u)\;+\;(\nabla_u T)(U, u)\;+\;R(U, u)u \;=\; 0,
\end{equation}
or, using the above notation, in the form
\begin{equation}\label{CCC}
U'' \;+\;T(U', u)\;+\; T\,'(U, u)\;+\;R(U, u)u \;=\; 0;
\end{equation}
notice that this equation, unlike equation (\ref{B}), involves the first derivative of $U$, not just the second. But we still have a linear second-order ordinary differential equation, and so the curvature and the torsion \emph{together} determine $U$ uniquely, once initial conditions are specified (though there is an interesting complication: we also need the derivative of the torsion along the autoparallel).

Again, if a physical spacetime theory allows us to compute the curvature and the torsion of three-dimensional space, we can solve equation (\ref{CCC}) for $U$ and compute $U'',$ the geodesic deviation. As before, we can then compare it with observations.

In order to proceed, we will have to extend the classical theory of submanifolds to the torsional case, to obtain relations between four-dimensional and three-dimensional\footnote{Actually, everything we have to say applies equally to $(n - 1)$-dimensional local spatial sections in an $n$-dimensional spacetime.} quantities. We now turn to the technicalities of that.

\addtocounter{section}{1}
\subsection* {\large{\textsf{3.2. Torsion and Submanifolds}}}
We begin with the torsion; then we move on to its derivative; and finally we treat the curvature. (We follow, and suitably generalise, the discussion in Section 3 of Chapter VII in \cite{kn:kobnom2}. Note that work along these lines has been done in the case of a specific form of torsion, corresponding to ``semi-symmetric'' linear connections: see \cite{kn:li}. The physics of semi-symmetric spacetime connections has been explored in \cite{kn:lehel1,kn:lehel2}.)

Let $M$ be a spacetime, that is, a smooth manifold endowed with a linear connection and a compatible metric of the appropriate signature, and let $\Sigma$ be a spacelike codimension 1 submanifold. Let $\nabla^*$ denote the covariant differentiation operator on $M$. Then for any vector fields $X$ and $Y$ on $\Sigma$, we can evaluate $\nabla^*_XY$ at an arbitrary point $x$ in $\Sigma$ as the sum of its tangential and normal components,
\begin{equation}\label{D}
\left(\nabla^*_XY\right)_x \;=\; \left(\nabla_XY\right)_x\;+\;\alpha_x\left(X,\,Y\right),
\end{equation}
where, initially, the two expressions on the right are no more than names for the respective tangential and normal components. Thus, in particular, $\alpha_x$ takes its values in the space of normal vectors to $\Sigma$ at $x$. This construction, valid separately at each $x$, defines a field over $\Sigma$, and with this understanding we drop the subscript $x$ henceforth (and likewise for $\nabla_XY$).

In (semi-)Riemannian geometry, $\alpha$ has a direct and important geometric meaning: it measures the rate at which the local unit normal field to $\Sigma$ varies as we move around\footnote{More precisely: if $X$ and $Y$ are unit vector fields on $\Sigma$, and $\xi$ is a unit local normal vector field to $\Sigma$, then $|\alpha(X,\,Y)|$ is the $Y$ component of the directional derivative of $\xi$ in the direction of $X$: see equation (\ref{T}) below. Therefore, if $e_i$ is a local basis of orthonormal vector fields on $\Sigma,$ then $\sum_i|\alpha(X,\,e_i)|e_i$ is the tangential part of the (spacetime) directional derivative of $\xi$ relative to $X$. Notice that, from a geometrical point of view, there is no reason to think that $\alpha(Y,\,X)$ should have any particular relation to $\alpha(X,\,Y)$. Generically, it will not, and, as we are about to see, that means that, generically, the torsion does not vanish.} in $\Sigma$. Therefore it is a measure of the way $\Sigma$ bends in $M$: it determines the ``extrinsic'' geometry of $\Sigma$. It can be computed if we know the spacetime covariant derivative operator and the local unit normal to $\Sigma$. In physical terms, this local unit normal field is known to us, because it is simply the field defined by the unit tangent vectors to the worldlines of a definite, fixed family of observers. (As usual, these observers can in principle be chosen freely, but in practice there is usually a distinguished choice, as for example in the case of FRW cosmologies.)

It turns out (see Appendix 1 below) that \emph{this interpretation of $\alpha$ remains exactly valid} when torsion is present (provided that the metric is compatible with the connection ---$\,$ that is, \emph{not} in theories with ``non-metricity''). Thus, even if the torsion does not vanish, $\alpha$ continues to describe the way $\Sigma$ is embedded in $M$ (and that is why it appears in the same way in the Gauss equation, equation (\ref{FFFF}) below, whether or not torsion is present). It also, as we are about to see, determines the ``extrinsic torsion''. We continue to call it the second fundamental form with respect to the given geometry of $M$ and the distinguished observer field (equivalently, the embedding of $\Sigma$) in $M$.

It is straightforward to show that $\nabla$ is in fact the covariant derivative operator for the pull-back of the linear connection to $\Sigma$. (It follows that $\alpha$, as a field on $\Sigma$, is bilinear over smooth functions on $\Sigma$.) One can also show that, because the spacetime metric is compatible with $\nabla^*$, it follows that the induced metric on $\Sigma$ is compatible with $\nabla$. Notice that if we know $\nabla^*$ and $\alpha$, then equation (\ref{D}) gives us $\nabla$ (which we need in order to compute the derivatives in the equation of geodesic deviation).

Extending $X$ and $Y$ locally to vector fields $X^*$ and $Y^*$ on $M$, the restriction of the commutator $[X^*,\,Y^*]$ to $\Sigma$ is tangential to $\Sigma$ and in fact coincides with $[X,\,Y]$. We then have
\begin{equation}\label{E}
\nabla^*_{X^*}Y^*\;-\; \nabla^*_{Y^*}X^*\;-\;[X^*,\,Y^*]\;=\; \nabla_{X}Y\;-\; \nabla_{Y}X\;-\;[X,\,Y]\;+\;\alpha(X,\,Y)\;-\;\alpha(Y,\,X).
\end{equation}
Denoting the torsion of $M$ by $T^*$, and that of $\Sigma$ by $T$, we then have (restricting to fields tangential to $\Sigma$) what we shall call the \emph{torsional Gauss-Codazzi equation}\footnote{The Gauss-Codazzi equations for curvature are traditionally stated as two separate equations, one tangential, one normal, but that is mainly for historical reasons and for convenience. The case of torsion is sufficiently simple that such a separation is unnecessary here.},
\begin{equation}\label{F}
T^*(X,\,Y)\;=\; T(X,\,Y)\;+\;\alpha(X,\,Y)\;-\;\alpha(Y,\,X).
\end{equation}
Given the spacetime torsion and the details of the embedding of $\Sigma$ in $M$, this equation allows us to find the torsion of $\Sigma$. Notice that we can do this without knowing the curvature of either $M$ or $\Sigma$.

A comparison with the more familiar Gauss equation for curvature (see below) leads us to some obvious definitions: $T$ is the \emph{intrinsic torsion}, and $\alpha(X,\,Y)\;-\;\alpha(Y,\,X)$ is the \emph{extrinsic torsion}.

The first term on the right of equation (\ref{F}) is tangential to $\Sigma$, while the second and third are normal to it; so if $T^* = 0,$ then, whatever $\alpha$ may be, $T$ too must vanish: if the torsion of the spacetime vanishes, so too must the torsion of any spatial section. (It also follows that $\alpha$ must be symmetric in that case, though in general it need not be.)

To put it another way: if we can establish that a spacelike section has non-zero torsion, then the full spacetime must likewise have non-zero torsion; this is true \emph{no matter what value $\alpha$ has}, because $T$ and $\alpha$ take their values in orthogonal spaces. That is, the statement does not depend on how the spacelike section is embedded.

This is remarkably different to the case of curvature, because it is of course perfectly possible to foliate (for example) Minkowski spacetime with curved spatial sections; spatial curvature does not imply spacetime curvature. This is the counterpart of the situation mentioned earlier, where we discussed the ``cancellation'' of the tangential component of spacetime curvature by the extrinsic curvature in FRW spacetimes with Euclidean spatial sections. In Minkowski spacetime we can use the extrinsic curvature to ``cancel'' the intrinsic curvature. As before, however, this is only possible because the spacetime has such a high degree of symmetry; but an approximate cancellation of this sort is still possible. Torsion is different: \emph{it cannot be ``cancelled'' by manipulating $\alpha$}, not even approximately.

While all this is of considerable interest, and we will return to it, we still have work to do: in order to use equation (\ref{C}) on $\Sigma$, we need to be able (as in the preceding Section) to compute the coefficients in the equation. These are given by the three-dimensional curvature tensor, and the derivative of the three-dimensional torsion.

Specifically: we assume that (as a result of solving field equations, and so on) that we know the \emph{spacetime} geometry, and we take it that we have chosen a family of observers, with the corresponding foliation by spacelike hypersurfaces, which determines a definite second fundamental form $\alpha$. Our objective is to use these data to determine the geometry of a spacelike hypersurface $\Sigma$, and so to compute (in terms of $u$ and $U$) the quantity $\nabla_u\left[T(U,\, u)\right]$ that occurs in the Jacobi equation. We proceed as follows.

First, fix a pair of vector fields $X$, $Y$ on $\Sigma$, and then define the $(1, 1)$ tensor $\nabla\left[T(X,\,Y)\right].$ It will in fact turn out to be much more convenient to work with the metrically equivalent $(0, 2)$ tensor $S_{X, \,Y}$ defined, for any pair of vector fields $W$ and $Z$ on $\Sigma$, by
\begin{equation}\label{FF}
S_{X,\, Y}(W,\,Z) = g(Z,\, \nabla_W\left[T(X,\, Y)\right]),
\end{equation}
where $g$ is the induced metric on $\Sigma.$

We will denote the similar object defined globally on $M$ (by using $\nabla^*$ instead of $\nabla$, $T^*$ instead of $T$, $g^*$, the spacetime metric, instead of $g$, and by using vector fields not necessarily tangential to $\Sigma$) by $S^*_{X, \,Y}$. (However, we are really only interested in the restriction of $S^*_{X, \,Y}$ to $\Sigma$, and we always evaluate it using vector fields which \emph{are} tangential.)

The equation we need is
\begin{equation}\label{FFF}
S^*_{X,\, Y}(W,\,Z)\;=\;S_{X,\, Y}(W,\,Z)\,-\,g^*\left(\alpha(W,\,Z),\;\;\alpha(X,\,Y)\,-\,\alpha(Y,\,X)\right).
\end{equation}

This can be obtained using the torsional Gauss-Codazzi equation (\ref{F}); for the derivation, see Appendix 1. In the now-familiar manner, if we are given, by (say) the Einstein-Cartan theory, the spacetime metric and torsion (and therefore the spacetime covariant derivative operator), we can compute (for any pair of vector fields $X$ and $Y$ tangent to $\Sigma$) the tensor $S^*_{X,\, Y}$, which is equivalent to $\nabla^*\left[T(X,\,Y)\right].$ Combining this with the details of the embedding of $\Sigma$ in $M$, we can compute $S_{X,\,Y}$ using equation (\ref{FFF}), and that gives us $\nabla\left[T(X,\,Y)\right].$ This can then be used in equation (\ref{C}).

Now we turn to the curvature. If (by the usual abuse of notation) we let $R^*$ and $R$ denote the $(0,\,4)$ versions of the four/three-dimensional curvature tensors respectively (meaning that $R^*(W,\,Z,\,X,\,Y)$ denotes $g^*(W,\,R^*(X,\,Y)Z)$, and similarly for $R$), then it can be shown (see Appendix 1 for a brief discussion) that, for vector fields $W, X, Y, Z$ tangential to $\Sigma$,
\begin{equation}\label{FFFF}
R^*(W,\,Z,\,X,\,Y)\;=\; R(W,\,Z,\,X,\,Y)\;+\;g^*(\alpha(X,\,Z),\,\alpha(W,\,Y))\;-\;g^*(\alpha(Y,\,Z),\,\alpha(W,\,X));
\end{equation}
this we can call the \emph{Gauss submanifold curvature equation}\footnote{See \cite{kn:jo1} for recent discussions of torsional ``Gauss-Codazzi equations'' in a rather different context.}. The sum of the two terms quadratic in $\alpha$ is what we have been calling the extrinsic curvature; let us give it a name, $E$, so that the terms in $\alpha$ are written as $E(W,\,Z,\,X,\,Y)$.

This equation has (for reasons explained earlier) precisely the same form as the semi-Riemannian Gauss submanifold curvature equation, but in fact it is quite different, because in general $\alpha$ is (as we saw above) not symmetric. The extrinsic curvature tensor $E$ therefore does not have some of the algebraic symmetries of the semi-Riemannian case, and this is appropriate because the curvature tensors likewise do not have those symmetries in the presence of torsion\footnote{We are thinking here primarily of the algebraic Bianchi identity. The usual antisymmetry under exchange of $W$ and $Z$, by contrast, still holds here, essentially because any metric connection on an $n$-dimensional manifold reduces to a connection on the bundle of orthnormal frames, so the curvature form takes its values in the Lie algebra of the special orthogonal group $SO(n)$; that is, in the algebra of $n \times n$ antisymmetric real matrices.}. In any case, all three objects belong to the relevant vector space of curvature tensors.

Equation (\ref{FFFF}) allows us to compute the curvature of $\Sigma$ given that of the spacetime, and now, combining this with our earlier discussion, we can solve equation (\ref{C}) for $X$, predict the three-dimensional geodesic deviation, and see whether it agrees with Gauss-style measurements.

We stress again that this procedure works without relying on any hypotheses regarding the worldlines of particles. Now we see that it works whether or not torsion is present. As we argued earlier, this is particularly important in the torsional case, because, in general, the worldlines of free particles in such theories do not probe the linear connection that really describes the actual geometry of spacetime.

The particular way in which torsion helps to govern geodesic deviation can lead to unfamiliar behaviour. To underline this point, consider the following scenario. Suppose that, in some torsional theory, we have a distribution of matter (and spin, in theories like the well-known Einstein-Cartan theory) such that both the curvature and torsion tensors are predicted to be negligible, in some specified spacelike section. Using a suitable extension of the tensor harmonics method described in \cite{kn:kumar}, we nevertheless find that the spatial section exhibits strong geodesic deviations. Should we conclude that our theory has been falsified?

By no means. For, looking at equation (\ref{CCC}), we see that the spatial geodesic deviation can be large even if both the curvature and the torsion are too small to be detected. This will happen if the reference autoparallel crosses what might be called a ``torsion escarpment'': a region in which there is a very steep but brief increase in the torsion, that is, a steep change in its value from one small quantity to another, \emph{still small}, quantity. For this will mean that the derivative term, $T\,'(X, u)$ will be large in that region, leading to a large value for $X''$. In the Einstein-Cartan theory, for example, this would arise immediately from a sudden but small increase in the spin density in the path of the reference autoparallel (because, in that theory, the relation between spin density and torsion is algebraic). \emph{Small torsion does not imply small geodesic deviations}.

No such effect can be attributed to the curvature, since its derivative does not occur in equation (\ref{CCC}). We would therefore have to conclude, under these circumstances, that torsion is present in three-dimensional space; and that means, as we saw when we discussed the torsional Gauss-Codazzi equation (\ref{F}), that we would have to conclude that torsion is present in the full spacetime.

In summary: torsion has the same standing as curvature, in the sense that it appears explicitly in the Jacobi equation (and also in another, more abstract sense explained in Appendix 2). But its geometry can be very different.

Thus far, we have been proceeding optimistically on the assumption that spatial torsion (or its derivatives) might be large enough to be directly observable. As is well known, however, that optimism would have been misplaced had we been discussing spatial \emph{curvature}. It is generally thought that spatial curvature ``inflates away'' in the early Universe, and it would not be surprising if the spatial torsion were ``inflated away'' with it. We will argue, however, that this does not mean the end of our hopes. For the intrinsic geometry of the spatial sections is not the full story here: we must also ask how the spatial sections bend into spacetime.

Clearly the theory of Inflation is directly relevant to our concerns here, and so we now turn to it.

\addtocounter{section}{1}
\section* {\large{\textsf{4. Torsion and Inflation}}}
Our objective in this Section is to discuss cosmic Inflation \cite{kn:inflationaris} and how it might work in the presence of torsion. For the sake of clarity, however, we first discuss Inflation in the zero-torsion case: the torsion tensor is set equal to zero until further notice.

\addtocounter{section}{1}
\subsection* {\large{\textsf{4.1. Inflation, Intrinsic Curvature, Extrinsic Curvature}}}
To the modern eye (though not to Gauss), perhaps the strangest aspect of Gauss' supposed investigations of spatial geometry is the fact that he failed to find anything of interest (which may well explain the fact that he wrote nothing about them, if indeed he considered the question). For we have long known that the geometry of our spacetime ---$\,$ let us take this to refer to a cosmological model of it ---$\,$ is far from trivial. \emph{Why then is space so different?} That is, why should non-Euclidean spatial geometry be so much harder to detect than non-Minkowskian spacetime geometry? Naively: if spacetime is curved, \emph{why is space Euclidean}?

For observations suggest \cite{kn:efstath,kn:suhail,kn:bengal} that the spatial sections of our Universe, as defined by distinguished observers (us), have, on average, Euclidean geometry (and topology, but we do not discuss this here). This statement is not intended to mean that detailed observations, after the manner of Gauss, have been made regarding this geometry: one is not reporting a failed effort to detect spatial geodesic deviation. Instead, what is meant is that the \emph{intrinsic} spatial curvature makes a negligible contribution to the Friedmann equation; the fictitious ``density'' parameter $\Omega_k$ discussed earlier is very small. (Of course, no direct astronomical observations point to non-zero spatial curvature, a fact that puzzled Schwarzschild as long ago as 1900 \cite{kn:sch1,kn:sch2}.)

This is very different to the overwhelming abundance of evidence for non-trivial \emph{spacetime} cosmological geometry.

That evidence comes to us, however, only indirectly: through the \emph{extrinsic} geometry of the distinguished spatial sections. To see this, we proceed as follows. Using a proper time coordinate $t$ such that the unit normal vector $\xi = \partial_t)$ we can put a cosmological spacetime metric into the form $-\,dt^2\,+\,a(t)^2\,f,$  where $f$ is an arbitrary time-independent three-dimensional spatial metric tensor (so that $g = a(t)^2\,f.$). Now since $\Sigma$ is of codimension one, we can set $\alpha\,=\,h\,\xi$, where $h$ is just an ordinary real bilinear form. We can show (see equations (\ref{TAU}) and (\ref{TTAU}), Appendix 1) that
\begin{equation}\label{FFZEROTOR}
h(X,\,Y)\,=\,{\dot{a}\over a}\,g(X,\,Y),
\end{equation}
where $X$ and $Y$ are arbitrary tangential fields to $\Sigma$, and the dot denotes a derivative with respect to proper time. We see very plainly that the Hubble parameter $\dot{a}/a,$ which henceforth we denote by H, is nothing but \emph{a measure of the second fundamental form}, of the way space bends into spacetime. (If we think in terms of the $(1, 1)$ version of $h$, then H is its eigenvalue.)

We see now from equation (\ref{FFFF}) that for \emph{any} spatial metric tensor of the form $g = a(t)^2\,f,$ the extrinsic curvature tensor is given by
\begin{equation}\label{ANY}
E(W,\,Z,\,X,\,Y)\,=\, \m{H}^2\,\left(g(Y,\,Z)\,g(W,\,X)\;-\;g(X,\,Z)\,g(W,\,Y)\right),
\end{equation}
where, here and henceforth, $W, X, Y, Z$ are all tangential to $\Sigma$.

Thus we see that the extrinsic curvature is essentially the square of the Hubble parameter, which certainly is measured to be non-zero, to the extent indeed that we can have disputes as to its precise value \cite{kn:wendy}. Given that the intrinsic curvature is negligible, the fact that the extrinsic curvature is large, together with the Gauss equation (\ref{FFFF}), means that the spacetime curvature cannot be negligible. So when we assert that cosmological spacetimes are known to be curved, what we really mean is that the second fundamental form (as manifested in the extrinsic curvature) of the distinguished foliation is \emph{large enough to be readily observed}.

So really we have \emph{two} observations to explain: the smallness of the intrinsic, \emph{and} the largeness of the extrinsic curvature in our Universe. Something has made the space we inhabit bend strongly into spacetime, while keeping its geometry Euclidean; a somewhat bizarre state of affairs, one that certainly calls for explanation.

In a way, the second problem, the largeness of extrinsic curvature, is more pressing than the first. For intrinsic curvature can be hard to discern: any Riemannian manifold will \emph{seem} to be devoid of interesting geometry if its length scale is large enough relative to us, and that simple observation is sharpened in accelerating spacetimes (like de Sitter spacetime, which plays a key role in our later discussion) by the fact that a given observer in such a spacetime has causal access to a rapidly shrinking portion of the spatial hypersurfaces \cite{kn:gibbons}. By contrast, the extrinsic curvature is, as we saw, readily detected and quantified.

The situation becomes still more puzzling when we consider how things stood at the earliest cosmological times. In many theories of the earliest Universe, the magnitudes of the second fundamental form and of the extrinsic curvature are thought to have been very \emph{small}. For example, in the ``No Boundary'' theory of Hartle and Hawking \cite{kn:lehners}, the earliest spatial section marks a transition from Riemannian to semi-Riemannian geometry, meaning that the second fundamental form $\alpha$ has to make a transition from having a positive square to having a negative one; this implies that $\alpha$ vanishes on that surface, and hence so does the extrinsic curvature. (One says that the earliest spatial section is a surface of time-symmetry, meaning invariance under the inversion of the normal vector $\xi$, which clearly implies the vanishing of the second fundamental form on that surface.) The same is true in Penrose's theory \cite{kn:penrose} of the origin of the Second Law of thermodynamics\footnote{A rough outline of a proof runs as follows. In Penrose's theory, the spacetime Weyl tensor is postulated to vanish along the earliest spatial section. Since the Weyl tensor of a three-dimensional space vanishes identically, the Weyl tensor of a spacelike-foliated four-dimensional spacetime is expressed entirely in terms of the extrinsic curvature (by a straightforward adaptation of the Gauss submanifold equation), and this forces the extrinsic curvature of the initial section to be zero.}.

But if the extrinsic curvature was very small at the earliest times, then the intrinsic curvature must have been very large: this follows from the fact that the energy density then was very large\footnote{Unless there was a large negative energy component superimposed on the usual positive energy density, a possibility which, for simplicity, we will not consider here.} (and, therefore, so was the relevant component of the spacetime curvature, see below), combined with the Gauss equation for submanifold curvature.

Thus we see that, at least in some models of the very early Universe, the situation was the diametric opposite of the one we see today: large intrinsic, and small extrinsic curvature. How has this strange reversal from then to the present to be explained?

There are several suggestions for explaining the strange imbalance of intrinsic and extrinsic curvatures in the present Universe (though most of them focus on the former). Perhaps the reason is to be found in the study of the spatial surface along which the Universe was ``created from nothing'': perhaps, contrary to intuition, its intrinsic curvature was always small, apart from quantum fluctuations, for essentially topological reasons \cite{kn:brett}. Perhaps there is a distinguished measure on the relevant state space, and maybe that measure favours a small intrinsic curvature \cite{kn:sean} (or maybe not \cite{kn:turok}).

The generally accepted explanation, however, is dynamic: the intrinsic curvature and the extrinsic curvature \emph{evolved} in such a way that the inequality is reversed.
This is (part of) the theory of Inflation \cite{kn:inflationaris}. In particular, Inflation predicts, at least in the context of GR, that current deviations from Euclidean spatial geometry should not just be small: they should be unobservably small. However, it would be wrong to think of Inflation as a device for suppressing all forms of curvature. On the contrary, as we will explain, it predicts (in the simplest case at least) that the extrinsic curvature should be \emph{as large as it can be} in the relevant spacetime. That is, it explains the fact, too often taken for granted, that the extrinsic curvature is \emph{not} unobservably small. One says that the intrinsic curvature has been ``inflated away''. But one should \emph{also} say that the extrinsic curvature has been ``inflated up''.

This is a crucial point. Inflation does not simply make the intrinsic curvature disappear. Instead, it is \emph{replaced}, in a continuous manner through the inflationary era, by the extrinsic curvature. One can picture this as an ``inflationary seesaw''. This is in contrast to (for example) ``open'' FRW spacetimes (with zero cosmological constant), in which the matter content is indefinitely diluted, and the intrinsic curvature does in fact completely disappear.

Because of this dual function of ``inflating away'' one form of curvature, while ``inflating up'' another, Inflation sets up a reasonable (but unusual) initial value problem on the reheating \cite{kn:reheat} spatial section, so that the geometry can continue to evolve beyond that time. Because the initial value problem has this specific form, the spacetime arrives at its current state, which is the strange one we have been discussing: one in which, from the point of view of the spatial sections, all of the relevant spacetime curvature resides in the extrinsic curvature, leaving (apparently) no trace in the intrinsic curvature.

In short, Inflation answers both of our questions, in a very satisfactory manner. Of course, Inflation solves many another problem: perhaps most importantly, it explains the observed almost scale-invariant spectra of temperature anisotropies in the CMB. Here however we will be satisfied if we can extend the inflationary explanation of the current state of the cosmic intrinsic and extrinsic curvatures to torsional generalisations of GR. We will \emph{not} concern ourselves with any other aspect of Inflation; particularly not with more vexed questions, such as whether Inflation explains the ``horizon problem'' and so on.

Let us now examine how all this works, more explicitly.

During the ``explosive expansion'' phase of Inflation, spacetime was indistinguishable from the exact de Sitter spacetime, and so it need not be described (as Inflation often is) by the spatially Euclidean version of that spacetime. (Recall \cite{kn:ellie} that this spacetime can be foliated by sections of positive, zero, and negative curvature, though only in the first case does one obtain coordinates covering the entire spacetime.) Clearly we should \emph{not} use the Euclidean version if we wish to understand how spatial curvature ``inflates away''. Let us use, therefore, the version with sections having a large positive intrinsic curvature; so the spatial sections are three-spheres or quotients of such by finite groups. Then the scale factor is $L\cosh(t/L),$ where $L$ is the length scale defined by the cosmological constant (which is fixed by the initial energy density) and $t$ is proper time, taken to satisfy $t\,\geq \,0$. (That is, we ignore the contracting part of the spacetime; $t\,=\,0$ is the ``beginning of time''.)

As an important aside here: one could argue that there is little justification for taking the earliest spatial sections to be extremely symmetric (perfectly locally spherical in our case here); we will discuss this troublesome question in detail, later. For the moment, we ask the reader to interpret (``global'') de Sitter spacetime as a simple model in which \emph{the intrinsic spatial curvature is not zero} initially. This is essential because, as stressed earlier, in the absence of extrinsic curvature (and negative energy densities) the intrinsic curvature \emph{must} be large initially, because the energy density of matter is so large. We think of the spherical geometry in this model as a placeholder for ``some initial spatial geometry which is far from being Euclidean''. The hope is that this will help us to understand how more complicated intrinsic curvatures ``inflate away''.

The $(0, 4)$ version of the intrinsic spatial curvature tensor is given by the familiar formula (see \cite{kn:kobnom1}, Chapter V)
\begin{equation}\label{INTRINSIC}
R(W,\,Z,\,X,\,Y)\,=\, L^{-2}\,\m{sech}^2(t/L)\left(g(Y,\,Z)\,g(W,\,X)\;-\;g(X,\,Z)\,g(W,\,Y)\right),
\end{equation}
where we have used $a(t) = L\cosh(t/L),$ and where $g$ is the spatial metric. Clearly this object does indeed ``inflate away'', in a manner governed by the hyperbolic secant function.

On the other hand, according to these observers, the extrinsic curvature of these spatial sections is non-zero for $t \,>\, 0$. (This can be understood as a consequence of the fact that the unit normal field is not a Killing field here; that is, of the fact that the spatial geometry is dynamic.) In fact, the scalar coefficient of the extrinsic curvature \emph{grows} as time passes, though it is bounded:
\begin{equation}\label{EXTRINSIC}
E(W,\,Z,\,X,\,Y)\,=\, L^{-2}\,\tanh^2(t/L)\left(g(Y,\,Z)\,g(W,\,X)\;-\;g(X,\,Z)\,g(W,\,Y)\right);
\end{equation}
see equation (\ref{ANY}), above.

The extrinsic curvature grows with time simply because the second fundamental form does so: from equation (\ref{FFZEROTOR}) we have in this case
\begin{equation}\label{SECONDFF}
h(X,\,Y)\,=\,L^{-1}\tanh(t/L)\,g(X,\,Y).
\end{equation}
This is the key observation here: the ``inflating away'' of the intrinsic curvature is not the whole story of Inflation: \emph{the second fundamental form ``inflates up''}.

We see that de Sitter spacetime gives an accurate account of the earliest Universe, as we described it earlier: the intrinsic curvature is at its maximum at $t\,=\,0,$ while the extrinsic curvature is zero at that time.

Now notice a simple but crucial point. As we can see explicitly, the intrinsic curvature does indeed ``inflate away'' throughout the inflationary era. But the spatial component of the \emph{spacetime} curvature \emph{does not}. For the spacetime is de Sitter spacetime, a maximally symmetric spacetime, with \emph{constant} curvature: that is, the (magnitude of the) sectional curvature, evaluated on any two-dimensional section through the spacetime tangent spaces, is always the same, it is equal to $1/L^2$. This implies that the full spacetime curvature tensor $R^*$ takes the same form as in equation (\ref{INTRINSIC}), so that, when it is evaluated on tangential fields $W, Z, X, Y,$ the result is
\begin{equation}\label{TOTAL1}
R^*(W,\,Z,\,X,\,Y)\,=\, L^{-2}\,\left(g(Y,\,Z)\,g(W,\,X)\;-\;g(X,\,Z)\,g(W,\,Y)\right).
\end{equation}
In physical terms, the constancy of the coefficient on the right side of (\ref{TOTAL1}) is essential. For the full spacetime curvature tensor is related, through the Einstein equation, to the energy density, which is a fixed multiple of $1/L^2$. The constancy of the coefficient therefore corresponds to the fact that the energy density, during the slow-roll era, is \emph{not} diluted by the inflationary expansion: of course, it persists, unchanging.

How can this constancy of the spacetime curvature be reconciled with the ``inflating away'' of the intrinsic curvature? The answer is supplied by equation (\ref{FFFF}): the extrinsic curvature must increase, in just such a manner as to compensate for the decrease in the intrinsic curvature. And so it does: the elementary relation $\m{sech}^2 (x)\,+\,\tanh^2 (x) \,=\, 1$ (for all $x$) shows that (\ref{INTRINSIC}) and (\ref{TOTAL1}) combine to \emph{enforce} (\ref{EXTRINSIC}). That is, the Gauss equation forces the coefficient in (\ref{EXTRINSIC}) to ``inflate up'' towards its asymptotic upper bound, $L^{-2}$, the curvature of de Sitter spacetime itself.

Thus indeed Inflation explains both the ``inflating away'' of the intrinsic curvature \emph{and} the equally important ``inflating up'' of the extrinsic curvature (or, equivalently, of the second fundamental form, equation (\ref{SECONDFF})).

Eventually, the ``explosive expansion'' era draws to a close. By that time, the replacement of intrinsic by extrinsic curvature is effectively complete, so that the spatial sections are effectively Euclidean; the energy density now becomes time-dependent, and this defines a new natural foliation of the spacetime, the one given approximately by the spatially Euclidean version of de Sitter spacetime (with a scale factor involving $\exp(t/L)$ instead of $\cosh(t/L)$). As mentioned earlier, this is the usual version used to describe the late stages of Inflation, as it evolves towards reheating. This sets up specific initial conditions for the subsequent evolution up to the present time; this evolution preserves the small intrinsic curvature and the large extrinsic curvature, that is, it gives rise to the ``extrinsic curvature-dominated state'' we observe today; which is what we were trying to explain.

We can formulate this argument in a more general way by recalling that the Friedmann equation is really just the Gauss submanifold equation applied to the ``$3 + 1$ decomposition'' of FRW spacetimes. This is seen in the derivation of the \emph{constraint equation} (\cite{kn:wald}, Chapter 10) which has to be satisfied in order for the spatial geometry to evolve consistently.

Now set $\alpha\,=\,h\,\xi$, where $\xi$ is the local unit normal to $\Sigma$, and where $h$ is an ordinary symmetric bilinear form. Let $\hat{h}$ be the linear transformation (on vector fields tangential to $\Sigma$) constructed from $h$ by using the three-dimensional metric $g$. Then if $\rho$ is the energy density according to the observers with worldlines having $\xi$ as tangent, the constraint equation states
\begin{equation}\label{REH}
{16\,\pi \,G\over c^4}\,\rho\,=\,\m{Scal} \,+\,\left(\m{Tr}\,\hat{h}\right)^2\,-\,\m{Tr}\,\hat{h}^2,
\end{equation}
where Tr denotes the trace map, where $\m{Scal}$ is the three-dimensional (intrinsic) scalar curvature, and where the currently observed cosmological constant has, to avoid confusion, been absorbed into $\rho$. This equation (which we will derive below in the torsional case, where it takes a similar form) is derived by combining the Einstein equation with the Gauss equation for submanifolds ---$\,$ that is why the right side takes this specific shape, in terms of the second fundamental form. Equation (\ref{REH}), expressed in coordinates such that $\xi\,=\,\partial_t$, is easily seen to be the Friedmann equation,
\begin{equation}\label{REHA}
{16\pi G\over c^4}\,\rho\,=\,\m{Scal}\,+\,6\,\m{H}^2.
\end{equation}

However, we should not allow this familiar equation to distract attention from the more fundamental equation (\ref{REH}), where $\hat{h}$ plays the central role, not the Hubble parameter. \emph{The Friedmann equation simply expresses the fact that the energy density of spacetime (together with the intrinsic scalar curvature) determines the way spatial sections bend into spacetime}. That is, $\rho$ and Scal determine $\hat{h}$ directly, the Hubble parameter only indirectly.

We now have a thoroughly geometric description of Inflation in the absence of torsion. Now let us apply these techniques to the possibility of a torsional version of Inflation. (See for example \cite{kn:barrow2,kn:barrow1,kn:bravo} for torsional cosmology. It is interesting to note that the history of torsional cosmology goes back surprisingly far, back, in fact, to the work of Friedmann and Schouten in 1924: see \cite{kn:lehel2}.)

\addtocounter{section}{1}
\subsection* {\large{\textsf{4.2. Torsional Inflation}}}
We have argued that, from a geometric point of view, Inflation in GR is a process that replaces intrinsic with extrinsic curvature, explaining the observed facts that spatial geodesic deviations have been reduced to unobservable levels, while the Hubble parameter has become large. This geometric formulation of Inflation has the great virtue of being almost completely independent of one's choice \cite{kn:inflationaris} of inflationary theory; that is, our discussion applies to any theory of the earliest cosmos that involves almost exponential expansion terminating in some kind of reheating.

This generality is very useful, in view of the current lack of certainty regarding the precise mechanism underlying Inflation. We can now contemplate ``torsional Inflation'' as being the result of the physics of some underlying form of matter (perhaps a ``torsional inflaton'', or some effect taking the place of the inflaton \cite{kn:without}) which carries both an energy density and a spin density. We will not need to be more precise.

We postulate that ``torsional Inflation'' should affect the torsion in a manner precisely analogous to the picture we painted above: that is, we suppose that both the intrinsic curvature \emph{and} the intrinsic torsion inflate away during the inflationary epoch, while the extrinsic curvature \emph{and} the ``extrinsic torsion'' have to ``inflate up'' in order to compensate. Let us see how that might work.

From equation (\ref{F}) we see that there are obvious definitions of ``intrinsic torsion'' and ``extrinsic torsion'': if we denote the latter by $T^E$,
\begin{equation}\label{TE}
T^E(X,\,Y)\,=\,\alpha(X,\,Y)\;-\;\alpha(Y,\,X),
\end{equation}
where as usual $X$ and $Y$ are tangential to $\Sigma$. So we will use this terminology henceforth. It is also clear from equations (\ref{FFF}) and (\ref{FFFF}) that a similar splitting is possible for the derivative of the torsion and, exactly as before, for the curvature.

Equation (\ref{TE}) implies that the extrinsic torsion is a multiple of the unit timelike normal $\xi$, the multiple being given by (twice) the antisymmetric part of the bilinear form $h$ we introduced above in the discussion of equation (\ref{FFZEROTOR}); so let us study $h$ in more detail.

When torsion is present, (\ref{FFZEROTOR}) is replaced by (see equations (\ref{TAU}) and (\ref{TTAU}) in Appendix 1 below)
\begin{equation}\label{EXT}
h(X,\,Y)\,=\,{1\over 2}\left(\mathcal{L}_{\xi}\,g^*\right)(X,\,Y)\,+\,g\left(\tau K^*(X,\,\xi),\,Y\right),
\end{equation}
where $\mathcal{L}_{\xi}\,g^*$ is the Lie derivative of $g^*$ with respect to $\xi$, where $K^*$ is the usual spacetime \emph{contortion tensor} \cite{kn:telebook,kn:mavro}, constructed (as a $(1, 2)$ tensor) in a simple way from the spacetime torsion tensor, and where $\tau$ denotes the tangential component (the point being that $K^*(X,\,\xi)$ is not itself tangential to $\Sigma$ in general).

It is important to note that the contortion tensor, in its $(0, 3)$ version as it appears here, is always antisymmetric in $X$ and $Y$, so the second term on the right is exactly the antisymmetric part of $h$: that is, essentially, the contortion evaluated on $\xi$ (and on tangent vectors to $\Sigma$) in the manner shown is (half of) the extrinsic torsion (see equation (\ref{TE})). If we define a two-form $\varkappa$ on $\Sigma$ by
\begin{equation}\label{KAPPA}
\varkappa(X,\,Y)\,=\,2\,g\left(\tau K^*(X,\,\xi),\,Y\right),
\end{equation}
where $X$ and $Y$ are tangential to $\Sigma$, then $\varkappa$ deserves to be named the extrinsic torsion form, because $T^E(X,\,Y) = \varkappa(X,\,Y)\xi$.

Now equation (\ref{EXT}) takes the simple form
\begin{equation}\label{EXTRO}
h\,=\,{1\over 2}\mathcal{L}_{\xi}\,g^*\,+\,{1\over 2}\varkappa,
\end{equation}
and so we see that the second fundamental form is a simple \emph{mixture} of the two quantities describing the extrinsic geometry of a spatial section: one related to the extrinsic Levi-Civita curvature, the other being the extrinsic torsion. These two objects are equally important in determining the extrinsic geometry, and they should be treated accordingly, that is, in the same way.

The extrinsic curvature still has the same overall form as in the semi-Riemannian case,
\begin{equation}\label{EXTRINSICCURVTOR}
E(W,\,Z,\,X,\,Y)\,=\, h(Y,\,Z)\,h(W,\,X)\;-\;h(X,\,Z)\,h(W,\,Y),
\end{equation}
with $h$ given by equation (\ref{EXT}) (though now of course the ordering of the vectors is important). Again we see that $h$ is really the basic object here.

We see from equation (\ref{EXTRO}) that the second fundamental form, that is, the measure of the bending of space into spacetime, has two aspects in the presence of torsion. The first term on the right in (\ref{EXTRO}) describes the tendency of the spatial section to bend as it evolves with time. (It is zero if the timelike unit normal is a metric Killing vector.) A more physical way of saying this is that this part of the second fundamental form is described by the Hubble parameter H.

The second term on the right in (\ref{EXTRO}) is quite different: it represents instead a kind of inherent tendency for the spatial section to \emph{twist} into spacetime, \emph{independently} of the way the section evolves with time. This twisting is due to the way the contortion tensor (in equation (\ref{KAPPA})) links space (represented by $X$) with time (represented by $\xi$); it gives the geometric meaning of torsion in the submanifold context. In torsional theories, space is in a sense more closely intertwined with spacetime than in GR: they are bound together whether there is temporal evolution or not.

We saw earlier (equation (\ref{REH})) that, in the zero-torsion case, the Friedmann equation is just a link between the energy density and the second fundamental form, $\hat{h}$. Before we can proceed, we need to understand what happens to (\ref{REH}) when torsion is introduced. Let us briefly investigate this.

For simplicity, we focus first on a definite torsional theory, the Einstein-Cartan theory \cite{kn:blaghehl}. The field equation here takes the same form as in GR,
\begin{equation}\label{EC}
\m{Ric}^*\,-\,{1\over 2}\,\m{Scal}^*\,g^*\,+\,\Lambda\,g^*\,=\,{8\pi G\over c^4}\,S^*,
\end{equation}
where the asterisk denotes a spacetime quantity, $\m{Ric}^*$ and $\m{Scal^*}$ are respectively the Ricci and scalar curvatures of the torsional connection, $\Lambda$ is a fundamental cosmological constant, and $S^*$ is the stress-energy-momentum tensor. Of course we must take care, since the Ricci curvature is not always symmetric when torsion is present, and nor is $S^*$.

Let $e_i,\,i = 1,\,2,\,3$ be an orthonormal basis of the tangent space to $\Sigma$ at any point, and convert it to an orthonormal basis of the spacetime tangent space by adding the timelike unit normal $\xi$. Then denoting timelike components by a zero subscript, we see that the zero-zero component of (\ref{EC}) takes the form
\begin{equation}\label{Z1}
\m{Ric}^*_{00}\,+\,{1\over 2}\,\m{Scal}^*\,=\,{8\pi G\over c^4}\,\rho;
\end{equation}
as before, to avoid confusion we have absorbed any fundamental cosmological constant into $\rho$. Now from its definition, the scalar curvature is, in this basis,
\begin{equation}\label{Z2}
\m{Scal}^*\,=\,-\,\sum_{i = 1}^3R^*_{0i0i}\,-\,\sum_{i = 1}^3R^*_{i0i0}\,+\,\sum_{i = 1}^3\sum_{j = 1}^3R^*_{ijij}.
\end{equation}
We now make use of the fact that the curvature components are antisymmetric in \emph{both} the first and the second pairs of indices; as was mentioned earlier, this is valid even in the presence of torsion \emph{provided} that the connection is compatible with the metric. Under these conditions, then, the first and second terms on the right are equal to each other and to $\m{Ric}^*_{00}$. The Gauss equation (\ref{FFFF}) gives us an expression for the last term on the right, and so we have now
\begin{equation}\label{Z3}
\m{Scal}^*\,=\, -\,2\,\m{Ric}^*_{00}\,+\,\sum_{i = 1}^3\sum_{j = 1}^3R_{ijij}\,+\,\sum_{i = 1}^3\sum_{j = 1}^3E_{ijij},
\end{equation}
where $\sum_{i = 1}^3\sum_{j = 1}^3R_{ijij}$ is the intrinsic (spatial) scalar curvature $\m{Scal}$ and where $E$ is the extrinsic curvature, given by (\ref{EXTRINSICCURVTOR}): this is where the second fundamental form enters the discussion.

In order to proceed, we need to pay attention to the fact that $h$ need no longer be symmetric. We therefore define the $(1, 1)$ version of $h$, which we earlier named $\hat{h}$, by
\begin{equation}\label{Z4}
g(\hat{h}(X),\,Y)\,=\,h(X,\,Y),
\end{equation}
for any tangential $X,\,Y$. (Note that, since $h$ is not symmetric in general, we are making a definite choice between two possible $(1, 1)$ tensors here.)

Inserting (\ref{EXTRINSICCURVTOR}) and (\ref{Z4}) into equation (\ref{Z3}), we find after some computation that
\begin{equation}\label{Z5}
\m{Scal}^*\,=\, -\,2\,\m{Ric}^*_{00}\,+\,\m{Scal}\,+\,\left(\m{Tr}\,\hat{h}\right)^2\,-\,\m{Tr}\,\left(\hat{h}^T\hat{h}\right),
\end{equation}
where $\m{Tr}$ denotes the trace map as before, and the final term involves the transpose of $\hat{h}.$ Inserting this into equation (\ref{Z1}) we obtain the somewhat familiar-looking relation
\begin{equation}\label{Z6}
{16\pi G\over c^4}\,\rho\,=\,\m{Scal}\,+\,\left(\m{Tr}\,\hat{h}\right)^2\,-\,\m{Tr}\,\left(\hat{h}^T\hat{h}\right),
\end{equation}
where the two terms involving $\m{Ric}^*_{00}$ have cancelled, so that the right side involves \emph{only} the intrinsic metric $g$ and the intrinsic torsional connection, together with $\hat{h}$. This fact is of course the basis of the initial-value formulation, which clearly works here just as it does in GR.

In summary: the generalised Friedmann equation here apparently differs very little from its GR counterpart, though of course we have to bear in mind that $h$ need not be symmetric. (Note particularly that the fact that the final term on the right in equation (\ref{Z6}) involves the transpose of $\hat{h}$ turns out to be important.)

We have gone through this derivation to underline the fact that the Hubble parameter \emph{only} enters the generalised Friedmann equation through the second fundamental form, in which however it ``mixes'' with the extrinsic torsion. We have reached this conclusion in the particular context of the Einstein-Cartan theory, but it is clear from the discussion that this conclusion would not have been different had we used some more complex torsional theory. Even in such a theory, the Hubble parameter appears in the generalised Friedmann equation because it lurks inside the second fundamental form ---$\,$ but the second fundamental form will also contain the extrinsic torsion, in \emph{every} torsional theory with no non-metricity, because equation (\ref{EXTRO}) is a purely geometric relation which applies to \emph{every} geometry with a connection and a compatible metric.

In short: the mixing of the Hubble parameter with the extrinsic torsion is generic for torsional theories with zero non-metricity.

The equations simplify considerably if we adopt a time coordinate $t$ such that $\xi\,=\,\partial_t$. Then as usual we write $g\,=\,a(t)^2\,f,$ where $f$ is a constant metric on $\Sigma,$ and then equation (\ref{EXTRO}) may be written in the form
\begin{equation}\label{EXTHAT}
\hat{h}\,=\,\m{H}\,I\,+\,{1\over 2}\,\hat{\varkappa},
\end{equation}
where $I$ is the identity transformation and $\hat{\varkappa}$ is the $(1, 1)$ tensor (on $\Sigma$) $\hat{\varkappa}\,=\,2\,\tau K^*(\,-\,,\,\xi);$ here $-$ denotes an open ``slot''. (See equation (\ref{KAPPA}).)

As explained earlier, $\varkappa$ is a (real) antisymmetric transformation in three dimensions, so its eigenvalues (which must be either zero or pure imaginary, in conjugate pairs) are $\kappa i,\,- \kappa i,\,0,$ where $\kappa$ is real and $i = \sqrt{- 1}.$ Thus in fact $\varkappa$ is described by a single real parameter, $\kappa$ (which we might call the \emph{extrinsic torsion parameter}), just as the extrinsic Levi-Civita curvature is described by its sole eigenvalue, the Hubble parameter.

We have now
\begin{equation}\label{TRANS}
\hat{h}^T\hat{h}\,=\,\left(\m{H}\,I\,-\,{1\over 2}\,\varkappa\right)\left(\m{H}\,I\,+\,{1\over 2}\,\varkappa\right)\,=\,\m{H}^2\,I\,-\,{1\over 4}\,\varkappa^2.
\end{equation}
Of course, $\varkappa$ has zero trace, so the second term on the right in equation (\ref{Z6}) is $9\,\m{H}^2$, while the trace of the right side of (\ref{TRANS}) is $3\,\m{H}^2\,+\,\kappa^2/2,$  and so we have finally
\begin{equation}\label{Z7}
{16\pi G\over c^4}\,\rho\,=\,\m{Scal}\,+\,6\,\m{H}^2\,-\,{1\over 2}\,\kappa^2.
\end{equation}
Bear in mind that $\m{Scal}$ here is the scalar curvature of the full torsional connection on $\Sigma$, so the intrinsic spatial torsion and its derivatives are hidden in this term. (That is why the equation does not take the form normally used in discussions of the Einstein-Cartan theory, which involves the Levi-Civita scalar curvature rather than Scal; but we need this form for our purposes here.)

Again we stress that this equation has been derived in the context of the Einstein-Cartan theory. A more general torsional theory will give rise to a more complex dependence on $\rho$. But, as we discussed earlier, $H$ and $\kappa$ will appear in the same combination even in such theories.

Of course, (\ref{Z7}) can be re-written as a straightforward modification of the usual Friedmann equation (\ref{REHA}), by extracting the torsional terms from Scal and forcibly combining them with the term involving $\kappa^2$, to construct an ``effective energy density'' for torsion. That may be useful for studying the solutions (for discussing the avoidance of cosmic singularities, for example), but such a procedure is geometrically meaningless and, we assert, deeply misleading. Intrinsic and extrinsic torsion are very different objects, which (we conjecture) can evolve in very different ways, and there is nothing fundamental to be gained by combining them.

Equation (\ref{Z7}) has an interesting and unexpected form: the square of the Hubble parameter and the squared extrinsic torsion parameter occur in it with opposite signs. This is due to the fact that equation (\ref{Z6}) involves $\hat{h}^T\hat{h}$ and not $\hat{h}^2.$ Notice that (in the Einstein-Cartan case) the positivity of the energy density imposes an upper bound on the extrinsic torsion parameter:
\begin{equation}\label{Z8}
\kappa^2\,<\,2\,\m{Scal}\,+\,12\,\m{H}^2.
\end{equation}
This would be replaced by a more intricate inequality in a more general torsional theory.

We argued earlier that $\alpha$ (equivalently, $\hat{h}$) should essentially vanish in the earliest Universe, and that argument continues to apply here without modification. It follows from equation (\ref{EXTHAT}) that \emph{both} H and $\kappa$ must likewise have been small at that time (since the two parts of the right side are just the symmetric and antisymmetric parts of $\hat{h}$). Thus we have the analogous situation to the one that we discussed in the GR context, where we stressed that the initial extrinsic curvature (that is, essentially, H) was small. Notice that, if the (relevant component of the) \emph{spacetime} torsion is large initially (as we naturally expect, and as we know is the case for the spacetime Levi-Civita curvature in GR) then equation (\ref{F}) indicates that the initial intrinsic torsion must have been large, like the initial intrinsic Levi-Civita curvature, again as expected. The strong formal analogy of equation (\ref{F}) with equation (\ref{FFFF}) certainly suggests that all this should work.

Let us begin by assuming that Inflation affects the metric in the way we explained above: the intrinsic Levi-Civita curvature ``inflates away'', and the extrinsic Levi-Civita curvature ``inflates up'' (that is, the symmetric part of the second fundamental form does so). Now, generically, we should not expect that the \emph{intrinsic} torsion and its derivatives will be unaffected by the sudden change in the metric geometry that is the distinguishing feature of Inflation. Rather, we might expect them to inflate away too.

That is, even if the intrinsic spatial torsion is large when Inflation begins, it might well ``track'' the intrinsic Levi-Civita curvature. If we further assume that the derivatives of the intrinsic torsion also ``inflate away''\footnote{If we do not assume this, then the possibility of ``torsional escarpments'', discussed earlier, arises.}, then, since the full intrinsic curvature tensor can be expressed as the sum of the intrinsic Levi-Civita curvature with expressions involving the intrinsic torsion and its derivatives, this would mean that all of the terms on the right side of equation (\ref{CC}), applied to the spatial section, have inflated away, leaving no geodesic deviation and producing Euclidean spatial sections at the time of reheating and subsequently. Our hopes of directly observing non-trivial spatial geometry will be dashed. (This would also mean that the intrinsic scalar curvature $\m{Scal}$ in equation (\ref{Z6}) should be set equal to zero at reheating and subsequently.)

It seems natural, however, to assume that, if the coupling of metric to torsion causes the intrinsic torsion to track the intrinsic Levi-Civita curvature, then it should cause \emph{the extrinsic torsion to track the extrinsic Levi-Civita curvature} during the inflationary epoch. This looks particularly natural if one examines equation (\ref{EXTHAT}): one expects the two terms to grow together, so that the \emph{entire} second fundamental form (and not just its symmetric part) grows during Inflation, just as it does in GR.

To take a concrete example: suppose that the ``tracking'' works in the simplest possible way, such that $H$ and $\kappa$ are approximately equal by the time of reheating, just as they are at the start of Inflation. Then (in the Einstein-Cartan theory) computing the apparent Hubble parameter (using equation (\ref{Z7})) at that time will result in an \emph{underestimate} of the actual value by about 4 percent. Even if the ``tracking'' takes a more complicated form, we always obtain an underestimate in this way. It would be very interesting to determine whether this property is generic for torsional theories.

We certainly do not claim that these simplistic observations, which in any case only take us up to reheating, explain the notorious ``tension'' between different computations and observations of the Hubble parameter \cite{kn:wendy}. That is a problem of considerable complexity and subtlety, with no simple solution \cite{kn:verde}. We would like to stress, however, that, among all possible modifications of GR, allowing non-zero torsion has a strong claim to be the most natural way to study possible disagreements between theoretical and observational values for the Hubble parameter. We say this because of the structure of equation (\ref{EXTRO}) itself: when torsion is present, the extrinsic torsion \emph{automatically} mixes with the Hubble parameter, in a natural, indeed inevitable manner. One might even claim that some kind of theoretical/observational Hubble anomaly is to be \emph{expected} in torsional theories.

This is really the point we wish to make here: cosmic (extrinsic) torsion is the natural starting-point for an attack on this major problem, if one decides that a modification of GR is required to handle it. (See \cite{kn:tortens} for an example of a serious attempt to use torsion to understand the Hubble tension.)

Any attempt to explore this further would involve selecting a specific torsional theory, something we have tried to avoid in the present work in order to retain maximal generality. If one does so, one will have to give an explicit account of the way in which the extrinsic torsion ``tracks'' the Hubble parameter; in particular, one will have to arrange for this to happen in a manner that always respects the inequality (\ref{Z8}) (or whatever replaces it in a more complex torsional theory).

Such an attempt would also have to take into account the following rather interesting and deep aspect of the problem.

We stressed that the above analysis works for \emph{any} three-dimensional metric of the form $g = a(t)^2\,f$; we need not require $f$ to be the metric of a space of time-independent constant curvature. If in fact $f$ is a metric of that kind, that is, if the spatial sections are isotropic around every point (as in the usual FRW models), then it is shown in \cite{kn:barrow2} that the (0, 3) spacetime torsion tensor is given by
\begin{equation}\label{BARROWA}
g^*(A,\,T^*(B,\,C))\,=\, \phi \Big(\,g^*(A,\,B)\,g^*(C,\,\xi)\,-\,g^*(A,\,C)\,g^*(B,\,\xi)\Big),
\end{equation}
where as usual the asterisk denotes a spacetime (not spatial) object, where $A,\,B,\,C$ are spacetime vectors (not necessarily tangent to $\Sigma$) and where $\phi$ is a function of cosmic time. The (0, 3) spacetime contortion in this special case is, with the same notation,
\begin{equation}\label{BARROWB}
g^*(A,\,K^*(B,\,C))\,=\, 2\phi \Big(\,g^*(\xi,\,A)\,g^*(B,\,C)\,-\,g^*(\xi,\,B)\,g^*(A,\,C)\Big).
\end{equation}
It is immediate that, if $A$ and $B$ are chosen to be tangential to $\Sigma,$ and $C$ is taken to be $\xi$, then the expression on the right in (\ref{BARROWB}) becomes zero; and this means (equation (\ref{KAPPA})) that the extrinsic torsion is zero at the beginning of Inflation, as we required. Unfortunately, however, it does not then ``inflate up'' ---$\,$ it remains zero at all times. This is perhaps less surprising if we note that the intrinsic torsion, contrary to our expectations, is also zero at the beginning and subsequently ---$\,$ take $A,\,B$ and $C$ to be tangential in equation (\ref{BARROWA}). There is no ``torsional inflation'' in this case, simply because there is nothing to ``inflate away''.

The problem is that the high degree of spatial symmetry \emph{assumed at the outset} here is not compatible with having non-zero intrinsic torsion (see \cite{kn:barrow2}). However, it is generally (not universally \cite{kn:penrose}) thought that the pre-inflationary Universe was \emph{not} isotropic, that it became so in the early stages of Inflation itself (see for example \cite{kn:pit} and the work it cites). So the initial spatial section being chosen here is too symmetric, it is already in the kind of state to which Inflation should \emph{lead}, not the kind in which it might be expected to originate. (Reference \cite{kn:barrow2} is concerned with torsional FRW cosmologies in general, not specifically with Inflation.)

In short, we ought really to have chosen the tensor $f$ to be non-isotropic. For then the expressions on the right in equations (\ref{BARROWA}) and (\ref{BARROWB}) no longer correctly represent the spacetime torsion and contortion; so that, when the spacetime torsion is evaluated on tangential vectors, the result is no longer zero. In fact, as we explained earlier, the intrinsic torsion should initially have been large, like the initial intrinsic Levi-Civita curvature. Then ``torsional Inflation'' becomes possible.

The problem now is that, in that case, one has work to do to show that even \emph{ordinary} Inflation actually occurs. For it has long been known (see for example \cite{kn:der,kn:trodden} for entry points to the large literature) that it is difficult to initiate Inflation if the pre-inflationary space was not \emph{already} very homogeneous and everywhere isotropic on super-Hubble scales.

That result, however, is based on assuming the exact validity of GR, as the authors of \cite{kn:trodden} point out. It is possible that, if the pre-inflationary spatial sections were sufficiently anisotropic as to be able to harbour a large intrinsic torsion, the resulting non-GR geometry might suffice to invalidate the argument of \cite{kn:trodden} and so permit ordinary Inflation to begin, even though the spatial geometry was not very homogeneous. The initial intrinsic torsion, having performed this service, might then ``inflate away'' along with the intrinsic Levi-Civita curvature, tracking the metric as we have proposed, while the extrinsic torsion, as a by-product, ``inflates up'' along with the extrinsic Levi-Civita curvature. If this could be shown, it would be a very satisfactory solution to the problem of inflationary initial conditions.

Clearly, a re-examination of \cite{kn:trodden}, but assuming a large initial intrinsic torsion, would be highly desirable. Note that other kinds of non-GR theories (see for example \cite{kn:genly}) are indeed known to behave very differently to GR, precisely in the context of inflationary ``isotropisation''. It is not unreasonable to hope that a large initial intrinsic torsion might lead to similar results.

Finally, we do not wish this discussion, which after all is concerned only with cosmology, to lead the reader to conclude that intrinsic torsion is negligible at the present time under all circumstances. Observations at the solar system level would need to be understood in the context of a torsional version of the Schwarzschild (or Kerr) solution, and there is no reason to think that either the intrinsic or the extrinsic torsion will be negligible in such a case. (The first term on the right in equation (\ref{EXTRO}) will vanish, since the spacetime is static, but the second will not in general.)

To summarise overall: torsional spatial geometry might be observable, either directly (on the solar system scale) or indirectly (at the cosmological level, through the way spatial sections are embedded in and twist through spacetime). Either way, we would be taken full circle back to Gauss in Hannover: perhaps his hopes can be realised, in an unexpected manner indeed.

\addtocounter{section}{1}
\section* {\large{\textsf{5. Taking Torsion Seriously}}}
In summary: equation (\ref{C}) has several interesting properties. The most obvious is that the curvature alone is \emph{not} sufficient to specify geodesic deviation: one needs also to know the torsion. Curvature and torsion are equally responsible for this fundamental geometric phenomenon. This means that, contrary to what is sometimes said, torsion is \emph{not} ``just another field'', like some kind of exotic matter field\footnote{There are objections to non-zero torsion based on attempts to understand quantum gravity by using effective field theory. If one believes that torsion is ``just another field'' then this will lead to a plethora of couplings to other fields, because one has no symmetries to constrain the couplings; nothing, for example, to prevent the inclusion of a ``massive torsion'' field. The point we have been stressing here, however, is precisely that torsion describes spacetime \emph{geometry}: it does not make sense to speak of ``massive torsion'' unless one can suggest a geometric interpretation for such a thing, a tall order indeed. Again, bear in mind that anything one says about torsion can be said with equal justice about curvature.}. Not only that: the presence of torsion is the default. If it is indeed completely absent both from spacetime and from space, then that certainly needs to be explained.

Inflation might well provide an explanation for the absence of torsion from three-dimensional space; but, if that is true, then it also strongly suggests that torsion has a role to play in determining how three-dimensional space bends into spacetime. In that guise it might well influence our understanding of the Hubble parameter, with which it naturally ``mixes''. The symmetry between the two halves of the right side of equation (\ref{EXTRO}) tells us that we have some explaining to do if it turns out that torsion plays no such role.

In short: we should take torsion seriously.

Regrettably, the literature abounds with work demonstrating that this advice (and more general advice \cite{kn:ellis} in the same vein) has not been heeded. For example, one is often informed that torsion can be abolished by means of a ``field redefinition'', which means that one completely ignores the fact that spacetime has a definite geometry, which cannot be changed simply by re-labelling the fields used to describe it. A similar geometric heedlessness is responsible for claims that GR is ``dynamically equivalent'' to a teleparallel theory or even to a theory with non-metricity (see \cite{kn:trinity} and the work it cites). In torsional and non-metricity theories there is more to spacetime geometry than the metric tensor; the resulting geometries are distinct from each other and from semi-Riemannian geometry; that (one would have thought) is the point. Once this is seen, it is clear that no such trinitarian ``equivalences'' are possible in any meaningful sense.

The crucial point here is this: GR, and its extensions we have discussed here, \emph{are really two theories in one}. First, they provide theories of the motion of massive particles (and of massless ones, as limiting cases): the worldlines are \emph{timelike} curves of extremal proper time. To determine these curves, spacetime theories have to control what might be called the ``temporal geometry'' of spacetime. This is the aspect of GR (as a special case) that has attracted most attention, because it is the aspect that immediately connected with observations; and because it is the aspect that is responsible for the characteristic theoretical achievements of spacetime theories, among which the abolition of the whole notion of ``gravitational force'' is the most important\footnote{Contrary to what is sometimes said, there is no call for talk of ``gravitational force'' in torsional theories, any more than in standard GR. In all of these theories, Newton's First Law can be formulated as the assertion that the worldlines of \emph{free}, massive particles are curves of extremal length (proper time). That already permits predictions of the trajectories of such particles. No such particle is being compelled to do anything that it would not do if left to its own devices, no Second Law need be invoked to explain its behaviour, and so the ``force'' of gravity can continue to rest in peace \cite{kn:synge}.}.

Thus, \emph{one} aspect of GR is that it is a theory of gravitation, replacing Newton's theory.

But the theories we have considered, including GR, are \emph{also}, as we have seen, theories of spatial geometry (and of the embedding of space in spacetime). This second aspect, which has nothing to do with ``gravitation'' in the Newtonian sense, has attracted far less attention, because until recently spatial geometry, and to some extent its embedding in spacetime, have been too difficult to probe observationally in detail. That may soon change.

This ``two-for-one'' aspect is best illustrated by the most basic non-Minkowskian spacetime of all, the exterior Schwarzschild geometry. The ``temporal geometry'' here is familiar to every student. But this spacetime also has spatial sections (relative to the distinguished Killing observers) with a specific geometry \cite{kn:rothman}. If Gauss had been able to survey the spatial geometry in the space around the Sun with sufficient precision, the angle sums of his triangles would not have been 180 degrees\footnote{We note in passing that by equation (\ref{TTAU}) below, these spatial sections have zero extrinsic curvature, and so a probe of the Schwarzschild spatial curvature is automatically a probe of the spatial part of the spacetime curvature. As mentioned at the end of the preceding Section, however, the analogous claim is no longer valid if torsion is present.}. They might not agree with the GR (Schwarzschild) prediction either, of course, and then the door to measurements of spatial and spacetime torsion would finally open.

Some of the greatest achievements of GR were made possible by the recognition, by Synge \cite{kn:synge}, Pirani \cite{kn:pirani}, and others, of the central role of the curvature tensor, particularly as it appears in the geodesic deviation equation (\ref{C}): the pioneers took curvature seriously. Today, as we are confronted with a vast variety of approaches to ``beyond GR'' physics, torsion is the obvious candidate for taking us to the correct theory. A focus on suitably generalised geodesic deviation, combined with a resolute determination to take spacetime geometry seriously, may well once again serve as an essential guide to the way forward.

\addtocounter{section}{1}
\section* {\large{\textsf{APPENDIX 1: Equations (\ref{FFF}) and (\ref{FFFF})}}}
Here we give a derivation of equation (\ref{FFF}), and a sketch of the (similar) derivation of equation (\ref{FFFF}). As before, we follow the notation of Sections 3 and 4 of Chapter VII of \cite{kn:kobnom2}, to which we refer the reader for still greater detail.

Using equation (\ref{F}), we can write
\begin{equation}\label{J}
\nabla^*_W\left[T^*(X,\, Y)\right]\,=\,\nabla^*_W\left[T(X,\,Y)\,+\,\alpha(X,\,Y)\,-\,\alpha(Y,\,X)\right].
\end{equation}
Using equation (\ref{D}) we obtain
\begin{equation}\label{K}
\nabla^*_W\left[T^*(X,\, Y)\right]\,=\,\nabla_W\left[T(X,\,Y)\right]\,+\,\alpha(W,\,T(X,\,Y))\,+\,\nabla^*_W\left[\alpha(X,\,Y)\,-\,\alpha(Y,\,X)\right].
\end{equation}
Now, since $\Sigma$ is codimension 1 in $M$, we can write
\begin{equation}\label{L}
\alpha(X,\,Y)\,=\,h(X,\,Y)\xi,
\end{equation}
where $\xi$ is a local field of unit normals to $\Sigma$, and where $h$ is a field of (scalar-valued) bilinear forms on $\Sigma$. We therefore have
\begin{equation}\label{M}
\nabla^*_W\left[\alpha(X,\,Y)\right] \,=\, W\left[h(X,\,Y)\right]\xi\,+\,h(X,\,Y)\nabla^*_W \xi.
\end{equation}
Of course, $\xi$ is not tangent to $\Sigma$, and so we must expect $\nabla^*_W \xi$ to have both tangential and normal components. We write
\begin{equation}\label{N}
\nabla^*_W \xi\,=\,-\,A_{\xi}(W) \,+\,D_W\xi,
\end{equation}
where $A_{\xi}(W)$ is tangential and $D_W\xi$ is normal. Note that $A_{\xi}$ measures the rate at which $\xi$ changes as one moves about in $\Sigma$: it is a measure of what we mean by ``extrinsic'' geometry. As we will see below, when the ``metricity'' condition is assumed, it is metrically equivalent to $\alpha$, so the latter is a measure of the extrinsic geometry; and this remains true in the presence of torsion.

One can follow the arguments in Section 3, Chapter VII of \cite{kn:kobnom2} and show that $A_{\xi}$ is a linear map from tangent fields to tangent fields, while $D_W$ is a covariant derivative operator for the normal bundle. Substituting equation (\ref{N}) into equation (\ref{M}), we find
\begin{equation}\label{O}
\nabla^*_W\left[\alpha(X,\,Y)\right] \,=\, W\left[h(X,\,Y)\right]\xi\,+\,h(X,\,Y)\left[-\,A_{\xi}(W) \,+\,D_W\xi\right].
\end{equation}
Substituting this, and the version with $X$ and $Y$ transposed, into equation (\ref{K}), we obtain
\begin{equation}\label{P}
\nabla^*_W\left[T^*(X,\, Y)\right]\,=\,\nabla_W\left[T(X,\,Y)\right]\,+\,\alpha(W,\,T(X,\,Y))\,+\,W\left[h(X,\,Y)\right]\xi\,+\,h(X,\,Y)\left[-\,A_{\xi}(W)\,+\,D_W\xi \right] \nonumber
\end{equation}
\begin{equation}\label{Q}
\,-\,W\left[h(Y,\,X)\right]\xi\,-\,h(Y,\,X)\left[-\,A_{\xi}(W) \,+\,D_W\xi\right].
\end{equation}
We are really interested in the tangential component\footnote{Partly for technical reasons (because later we will take the inner product with a tangential vector field, and this eliminates the normal component anyway), but mainly because the quantity we are seeking, $\nabla_W\left[T(X,\,Y)\right],$ is of course tangential.} of this, which we obtain simply by dropping all terms involving $\alpha$ and $\xi$ directly, and $D_W\xi.$ So we have, if $\tau$ denotes taking the tangential component,
\begin{equation}\label{R}
\tau\nabla^*_W\left[T^*(X,\, Y)\right]\,=\,\nabla_W\left[T(X,\,Y)\right]\,-\,\left[h(X,\,Y)\,-\,h(Y,\,X)\right]A_{\xi}(W).
\end{equation}
Now let $Z$ be a vector field tangential to $\Sigma$, and recall that $g^*$ is the metric tensor of $M$. We have of course $g^*(\xi,\,Z)\,=\,0.$ Applying $\nabla^*_W$ to this equation and using our standing assumption that $\nabla^*$ is compatible with the metric (so our conclusions from this point would not be valid in a theory with ``non-metricity''), we have
\begin{equation}\label{S}
g^*(-\,A_{\xi}(W)\,+\,D_u\xi,\,Z)\,+\,g^*(\xi,\,\nabla_WZ\,+\,\alpha(W,\,Z))\,=\,0.
\end{equation}
Since $Z$ is perpendicular to $D_u\xi$ and $\nabla_WZ$ is perpendicular to $\xi,$ this gives us
\begin{equation}\label{T}
g(A_{\xi}(W),\,Z)\,=\,g^*(\xi,\,\alpha(W,\,Z)),
\end{equation}
where we see $g$ on the left because here $g^*$ is being evaluated only on tangential vector fields. Equation (\ref{T}) shows that, as claimed above, $\alpha$ can be regarded as a version of $A_{\xi}$, and therefore describes the extrinsic geometry in the usual manner, in any torsional theory with a metric compatible with the connection.

We digress slightly here to point out that equation (\ref{T}) can be used to derive an explicit expression for the second fundamental form, as follows. Combining this equation with the definition of $A_{\xi}$ and with equation (\ref{L}), we have, for tangential $X$ and $Y$,
\begin{equation}\label{TT}
h(X,\,Y)\,=\,g^*\left(\nabla_X^*\,\xi,\,Y\right).
\end{equation}
Let $e_i$ be a set of local basis fields tangential to $\Sigma$, and construct a set of spacetime basis fields by adjoining $\xi$; the index corresponding to the latter will be 0. Taking components relative to these basis vectors, one finds
\begin{equation}\label{TTT}
h_{ij}\,=\,g_{jk}\,\Gamma^{*k}_{i\,0},
\end{equation}
where the $\Gamma^{*k}_{i\, 0}$ are components of the spacetime linear connection, all indices apart from $0$ run from 1 to 3, and a summation is implied on $k$. Now set
\begin{equation}\label{TTTT}
\Gamma^{*k}_{i\,0}\,=\,\Delta^{*k}_{i\,0}\,+\,K^{*k}_{i\, 0},
\end{equation}
where the $\Delta^{*k}_{i\,0}$ are (some of) the components of the spacetime semi-Riemannian connection generated by $g^*$ (the Christoffel symbols), and where the $K^{*k}_{i\,0}$ are components of the spacetime contortion tensor $K^*$. A straightforward calculation shows now that equation (\ref{TTT}) is just
\begin{equation}\label{TAU}
h_{ij}\,=\,{1\over 2}\,\xi\,g_{ij}\,+\,g_{jk}\,K^{*k}_{i\,0},
\end{equation}
or
\begin{equation}\label{TTAU}
h(e_i,\,e_j)\,=\,{1\over 2}\left(\mathcal{L}_{\xi}\,g^*\right)(e_i,\,e_j)\,+\,g\left(\tau K^*(e_i,\,\xi),\,e_j\right),
\end{equation}
and this is essentially equation (\ref{EXT}) above.

Again we see that $h$ must be symmetric in the absence of torsion, but need not be so in its presence. We also see that, if $\xi$ is a timelike metric Killing vector, $h$ need not vanish, and we have an explicit expression for it in terms of the spatial metric and the spacetime contortion; we can express the extrinsic curvature in those terms if required.

To resume the discussion of equation (\ref{R}): we have
\begin{equation}\label{U}
g^*(Z,\,\nabla^*_W\left[T^*(X,\, Y)\right])\,=\,g(Z,\,\nabla_W\left[T(X,\,Y)\right])\,-\,h(X,\,Y)g(Z,\,A_{\xi}(W)) \,+\,h(Y,\,X)g(Z,\,A_{\xi}(W)).
\end{equation}
Notice that the introduction of the metric at this point automatically selects the tangential components. This is modelled on the discussion in Section 4 of Chapter VII of \cite{kn:kobnom2}, where the transition is made from the $(1,3)$ version of the curvature tensor to the $(0,4)$ version, to obtain the standard way of stating the Gauss equation for submanifold curvature.

Using equation (\ref{T}), we have
\begin{equation}\label{V}
g^*(Z,\,\nabla^*_u\left[T^*(X,\, Y)\right])\,=\,g(Z,\,\nabla_W\left[T(X,\,Y)\right])\,-\,g^*(\xi,\,\alpha(W,\,Z))\left[h(X,\,Y)\,-\,h(Y,\,X)\right].
\end{equation}
The definition of $h$ now gives us finally
\begin{equation}\label{W}
g^*(Z,\,\nabla^*_W\left[T^*(X,\, Y)\right])\,=\,g(Z,\,\nabla_W\left[T(X,\,Y)\right])\,-\,g^*(\alpha(W,\,Z),\,\alpha(X,\,Y)\,-\,\alpha(Y,\,X)),
\end{equation}
and this is equation (\ref{FFF}).

Equation (\ref{FFFF}) is derived by employing the same techniques as in the above discussion; as the proof is rather similar to the one given in Section 4 of Chapter VII of \cite{kn:kobnom2} (one must however take care to preserve orderings), we do not give it in detail. As in the discussion given here, the torsion tensor appears at various points, but all of its explicit appearances are removed at the step where one extracts the tangential component.

\addtocounter{section}{1}
\section* {\large{\textsf{APPENDIX 2: Torsion as a Kind of Curvature}}}
We saw that the equation of geodesic deviation (\ref{C}) shows that torsion and curvature are, in some sense, objects of the same status (though we have also seen that they can have quite distinct implications for space and spacetime geometry). Here we would like to draw attention to a way of looking at torsional geometry that allows us to formulate this ``similarity of status'' in an exact way (and thus to understand why torsion is so important). The reference here is Chapter III, Section 3 of \cite{kn:kobnom1}; we use the terminology of that work (so that, for example, a ``connection'' is a Lie algebra-valued one-form on a principal fibre bundle, and so on).

Tangent spaces to a differentiable manifold $M$ are normally thought of as vector spaces, but it is possible to think of them as manifolds in their own right, with the tangent vectors being regarded as points in that manifold. Let $A_x$ be a tangent space at $x \in M$ regarded in this way, and let $V_x$ be the tangent space regarded as a vector space, but therefore also as an Abelian group. Then $V_x$ acts as a group of motions on $A_x$ in the obvious way, and this action is transitive, so every element of $A_x$ can be expressed in the form $p \,+ \,ua,$ where $p \in A_x$, $u$ is a basis (interpreted as a linear isomorphism from $\bbr^n$, dim$(M) = n$, to $V_x$) for $V_x$, and $a \in \bbr^n$. The pair $(u,\,p)$ is called (in \cite{kn:kobnom1}) an \emph{affine frame} at $x$.

The group $A(n,\,\bbr)$, the semi-direct product of the general linear group $GL(n,\,\bbr)$ with $\bbr^n$ (regarded again as a group), has a natural free action to the right on the set of all affine frames over $M$, $A(M)$, and so the latter becomes the bundle of affine frames, a principal fibre bundle over $M$.

Let $F(M)$ be the usual bundle of frames over $M$, let $\omega$ be a connection on $F(M)$ (corresponding to what we normally call a linear connection), and let $\theta$ be the canonical one-form on $F(M)$ (see Chapter III, Section 2 of \cite{kn:kobnom1}). Since $\theta$ takes its values in $\bbr^n$, we can construct $\omega \,+\, \theta$ as a one-form on $F(M)$, taking its values in the Lie algebra of $A(n,\,\bbr)$ (the addition being done in that algebra).

Now we can use the natural injection $\gamma \,:\, F(M)\,\rightarrow \, A(M)$ to define a one-form $\tilde{\omega}$ on $F(M)$ by
\begin{equation}\label{X}
\gamma^*\tilde{\omega}\,=\,\omega \,+\, \theta,
\end{equation}
where $\gamma^*$ denotes the pull-back, and one can show that this $\tilde{\omega}$ is in fact a connection on $A(M)$. Each linear connection defines such an ``affine connection'' in this natural way.

But now something remarkable happens when we compute the curvature two-form, $\tilde{\Omega}$, of this connection: it is just
\begin{equation}\label{Y}
\gamma^*\tilde{\Omega}\,=\,\Omega \,+\, \Theta,
\end{equation}
where $\Omega$ is the usual curvature two-form, and \emph{where $\Theta$ is the torsion two-form on} $F(M)$.

In this way of formulating differential geometry, then, torsion and curvature are literally just two different components of the \emph{same} object, the affine curvature form. Torsion really is a kind of (that is, a component of) curvature in ``affine'' geometry. The fact that both curvature and torsion are needed to describe geodesic deviation is no longer a mystery; equation (\ref{C}) can be interpreted now as saying that geodesic deviation requires the full affine curvature for its description.

We close by noting that, when a metric is present, the general linear group can be replaced here by the orthogonal group, the bundle of affine frames reduces to the bundle of orthonormal affine frames, the connection (\emph{if} it is compatible with the metric) becomes a connection on that bundle, and so the structure group of the bundle becomes the $n$-dimensional version of the Poincar\'{e} group. In this limited sense, torsional theories are ``gauge theories of the Poincar\'{e} group''. This statement is useful, if only as inspiration for novel ways of thinking about gravitation; but the fact remains that torsion occurs in the geodesic deviation equation (\ref{C}), while gauge fields do not. Geodesic deviation is sensitive to both rotations and translations in ``tangent manifolds'', but not to gauge transformations, and this is a fundamental distinction, more essential even than the fact that gravity and gauge fields are governed by field equations of very different kinds.

\addtocounter{section}{1}
\section*{\large{\textsf{Acknowledgement}}}
The author is grateful to Dr. Soon Wanmei for helpful comments.

\end{document}